\documentclass[aps,prl,twocolumn,amsmath,amssymb,showpacs,superscriptaddress,notitlepage,longbibliography]{revtex4-1}
\usepackage[colorlinks=true,linkcolor=blue,anchorcolor=red,citecolor=blue, urlcolor=blue]{hyperref}
\usepackage{bm}
\usepackage{graphicx}
\usepackage{color}

\begin{document}

\title{Decays of Majorana or Andreev oscillations induced by steplike spin-orbit coupling}

\author{Zhan Cao}
\affiliation{Shenzhen Institute for Quantum Science and Engineering and Department of Physics, Southern University of Science and Technology, Shenzhen 518055, China}
\affiliation{School of Physics, Southeast University, Nanjing 211189, China}
\affiliation{Peng Cheng Laboratory, Shenzhen 518055, China}
\affiliation{Shenzhen Key Laboratory of Quantum Science and Engineering, Shenzhen 518055, China}

\author{Hao Zhang}
\affiliation{State Key Laboratory of Low Dimensional Quantum Physics, Department of Physics, Tsinghua University, Beijing 100084, China}
\affiliation{Beijing Academy of Quantum Information Sciences, Beijing 100193, China}

\author{Hai-Feng L\"{u}}
\affiliation{School of Physics, University of Electronic Science and Technology of China, Chengdu 610054, China}

\author{Wan-Xiu He}
\affiliation{Center for Interdisciplinary Studies and Key Laboratory for Magnetism and Magnetic Materials of the Ministry of Education, Lanzhou University, Lanzhou 730000, China}

\author{Hai-Zhou Lu}
\email{Corresponding author: luhaizhou@gmail.com}
\affiliation{Shenzhen Institute for Quantum Science and Engineering and Department of Physics, Southern University of Science and Technology, Shenzhen 518055, China}
\affiliation{Peng Cheng Laboratory, Shenzhen 518055, China}
\affiliation{Shenzhen Key Laboratory of Quantum Science and Engineering, Shenzhen 518055, China}

\author{X. C. Xie}
\affiliation{International Center for Quantum Materials, School of Physics, Peking University, Beijing 100871, China}
\affiliation{Beijing Academy of Quantum Information Sciences, Beijing 100193, China}
\affiliation{CAS Center for Excellence in Topological Quantum Computation, University of Chinese Academy of Sciences, Beijing 100190, China}

\begin{abstract}
The Majorana zero mode in the semiconductor-superconductor nanowire is one of the promising candidates for topological quantum computing. Recently, in islands of nanowires, subgap-state energies have been experimentally observed to oscillate as a function of the magnetic field, showing a signature of overlapped Majorana bound states. However, the oscillation amplitude either dies away after an overshoot or decays, sharply opposite to the theoretically predicted enhanced oscillations for Majorana bound states. We reveal that a steplike distribution of spin-orbit coupling in realistic devices can induce the decaying Majorana oscillations, resulting from the coupling-induced energy repulsion between the quasiparticle spectra on the two sides of the step. This steplike spin-orbit coupling can also lead to decaying oscillations in the spectrum of the Andreev bound states. For Coulomb-blockade peaks mediated by the Majorana bound states, the peak spacings have been predicted to correlate with peak heights by a $\pi/2$ phase shift, which was ambiguous in recent experiments and may be explained by the steplike spin-orbit coupling. Our work will inspire more works to reexamine effects of the nonuniform spin-orbit coupling, which is generally present in experimental devices.
\end{abstract}

\maketitle

Identifying and engineering Majorana bound states \cite{alicea2012new,leijnse2012introduction,beenakker2013search,stanescu2013majorana,aguado2017majorana} for topological quantum computing \cite{kitaev2003fault,nayak2008non,sarma2015majorana} remains a challenge. Among various candidates, the semiconductor-superconductor nanowires \cite{lutchyn2010majorana,oreg2010helical} have received considerable attention \cite{mourik2012signatures,deng2012anomalous,das2012zero,finck2013anomalous,churchill2013superconductor,deng2016majorana,chen2017experimental,suominen2017zero,nichele2017scaling,zhang2018quantized,
gul2018ballistic,sestoft2018engineering,vaitiekenas2018effective,deng2018nonlocality,de2018electric,bommer2018spin} due to their high tunability \cite{lutchyn2018majorana}. The Majorana bound states always come in a pair and are localized at the two ends of the wire. They are supposed to have zero energy, but in realistic nanowires within a few micrometers, the Majorana bound states hybridize. The hybridization energy $E_0$ is predicted to oscillate as a function of the Zeeman energy, chemical potential, or wire length \cite{prada2012transport,sarma2012splitting,rainis2013towards}, dubbed Majorana oscillations. Recent experiments in islands of nanowire find that $E_0$ oscillates with increasing magnetic field: the oscillation amplitude either dies away after an overshoot \cite{albrecht2016exponential,sherman2017normal,albrecht2017transport,vaitiekenas2018selective,o2018hybridization} or decays meanwhile the oscillation period in magnetic field increases \cite{albrecht2016exponential,o2018hybridization,shen2018parity}. However, these behaviors are sharply opposite to the theories for the Majorana bound states \cite{sarma2012splitting}, which predict an enhanced oscillation amplitude and period. Several theoretical studies \cite{chiu2017conductance,dmytruk2018suppression,fleckenstein2018decaying} have tried to address this discrepancy, but are partially successful, e.g., assumed multiple subbands and temperatures higher than those in the experiments \cite{chiu2017conductance} or found that the oscillation period decreases with increasing magnetic field  \cite{dmytruk2018suppression,fleckenstein2018decaying}. This discrepancy has raised the concerns on the conclusive identification of Majorana bound states, and has even endangered the scheme of Majorana qubits based on the nanowires \cite{plugge2017majorana,karzig2017scalable}.

\begin{figure}[b!]
\centering
\includegraphics[width=0.85\columnwidth]{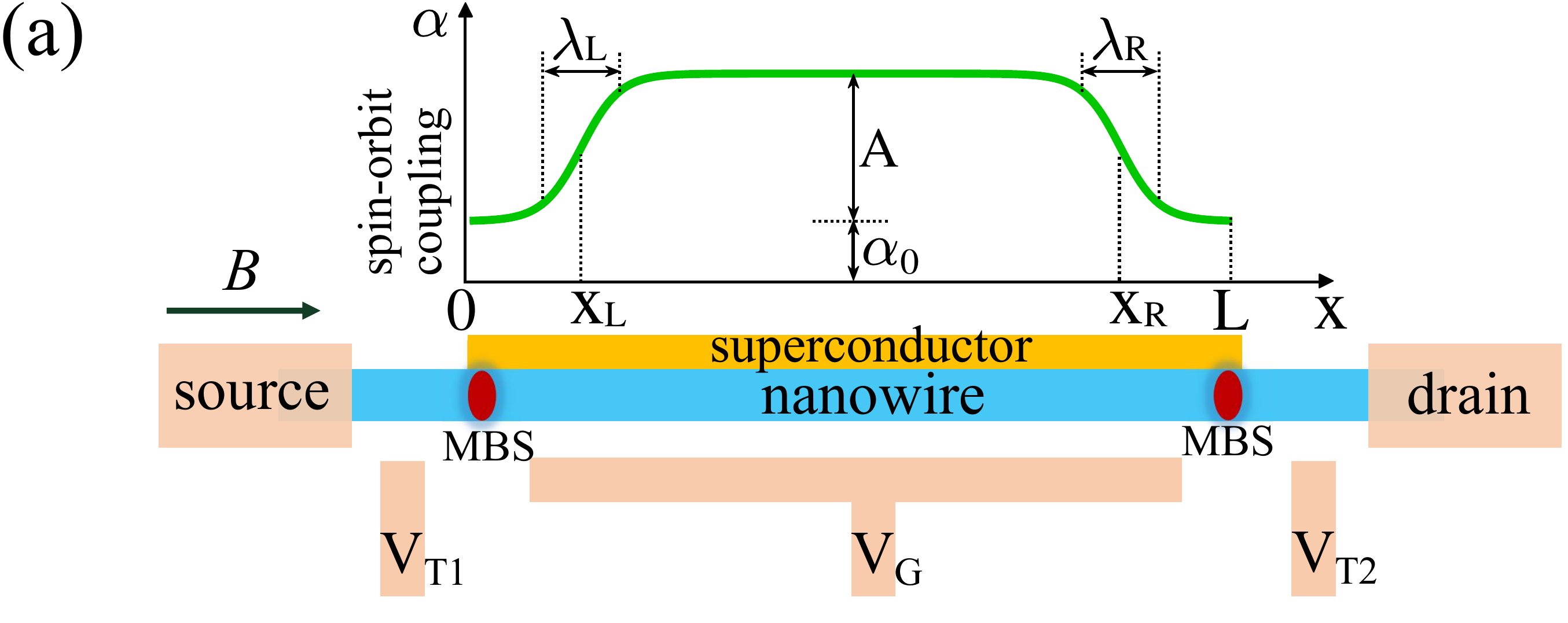}
\includegraphics[width=0.9\columnwidth]{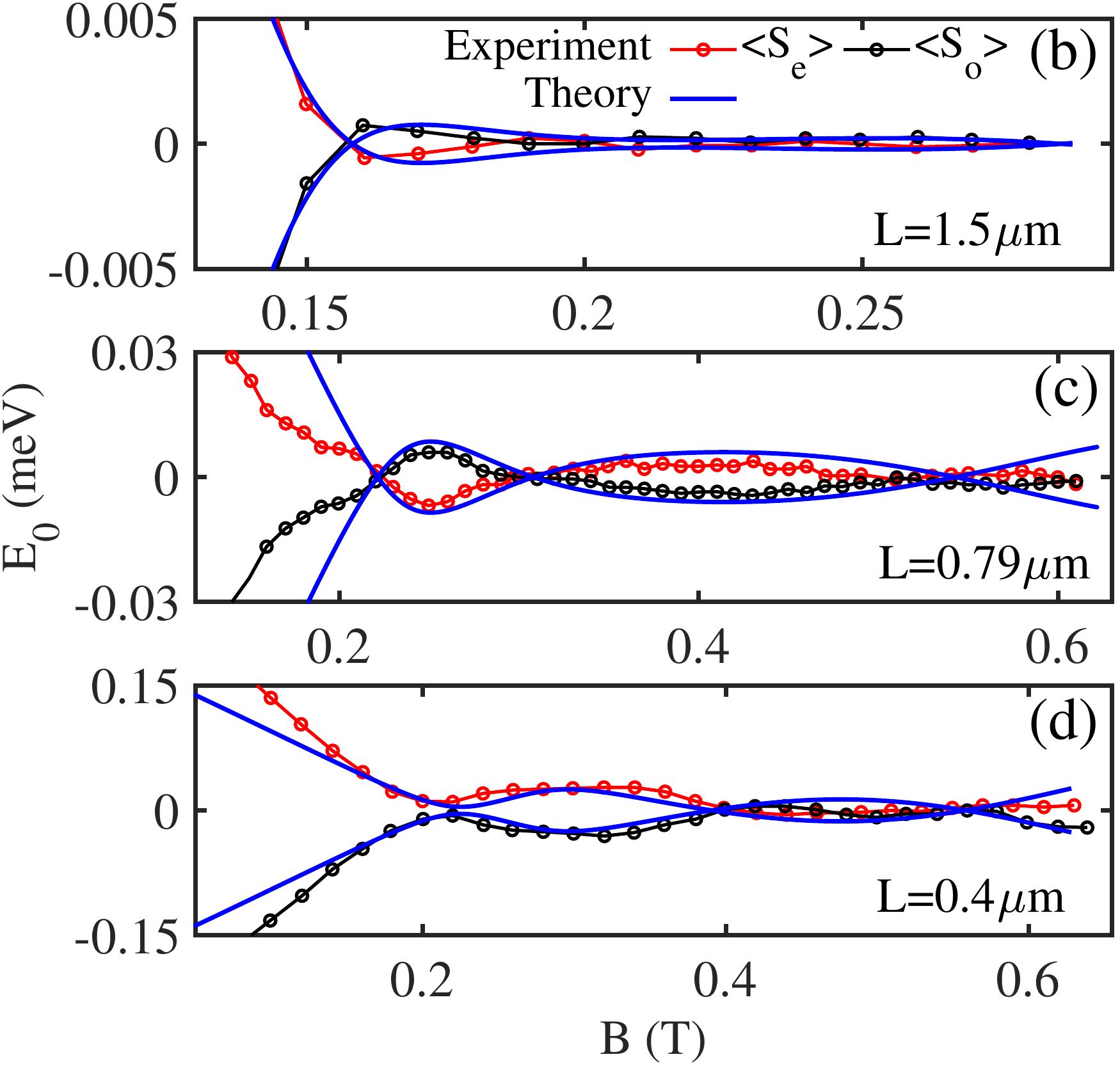}
\caption{(a) Schematic of the semiconductor-superconductor nanowire island \cite{albrecht2016exponential,sherman2017normal,albrecht2017transport,vaitiekenas2018selective,o2018hybridization,shen2018parity}, its two ends may host a pair of Majorana bound states (MBSs). [(b)-(d)] The red and black curves are adapted from Ref.~\cite{albrecht2016exponential}. The MBSs can hybridize. The hybridization energy $E_0$ in the experiments oscillates with decaying amplitude and increasing period as a function of the magnetic field $B$. However, opposite to the experiments, Majorana theory predicts that $E_0$ oscillates with increasing amplitude as a function of the $B$-induced Zeeman energy $V_Z$ \cite{sarma2012splitting}. By considering the steplike spin-orbit coupling $\alpha(x)$ in Fig.~\ref{Fig:WhyDecay}(a) (see the parameters listed in Sec.~SI of Ref.~\cite{Supp}), we find that the oscillation patterns of $E_0$, both the decay in amplitude and increase in period, can be well captured by the blue curves. See Fig.~\ref{Fig:Phase} for the relations between $\langle S_{e/o}\rangle$ and $E_0$. }\label{Fig:Decay}
\end{figure}

In this Letter, we reveal that the oscillation patterns in the experiments \cite{albrecht2016exponential,sherman2017normal,albrecht2017transport,vaitiekenas2018selective,o2018hybridization,shen2018parity}, including both the decay in amplitude and increase in period, can be well captured [Figs.~\ref{Fig:Decay}(b)-(d)] by a simple, but realistic assumption: spin-orbit coupling strength along the nanowire has a steplike distribution [see the green curve in Fig.~\ref{Fig:Decay}(a)]. The steplike spin-orbit coupling is reasonable because the gates apply a nonuniform electrostatic potential and spin-orbit coupling depends on the electrostatic fields perpendicular to the nanowire \cite{winkler2003spin,sanchez2006fano,sanchez2008strongly,glazov2011theory,sadreev2013effect,modugno2017macroscopic,klinovaja2015fermionic,dolcini2018magnetic}. Moreover, the presence of the superconductor can greatly modify the electrostatic field in the nanowire due to screening effect and work-function mismatch between the superconductor and semiconductor \cite{bommer2018spin}. Thus the spin-orbit coupling is well expected to be nonuniform from the nanowire covered with superconductor to the part (tunnel barrier region) without the superconductor. Additionally, we find that these decaying oscillations caused by the steplike spin-orbit coupling also exist in the energy spectrum of Andreev bound states [Figs.~\ref{Fig:Andreev}(d) and \ref{Fig:Andreev}(e)]. To distinguish Majorana from Andreev bound states, a recent theory \cite{hansen2018probing} predicted a $\pi/2$ phase shift between the spacings and heights of the Coulomb-blockade peaks mediated by the Majorana bound states in nanowire islands [Fig.~\ref{Fig:Phase}(b)]. The $\pi/2$ phase shift has been observed in Ref.~\cite{o2018hybridization}, but not in Ref.~\cite{shen2018parity}, which may be explained by considering the steplike spin-orbit coupling [Figs.~\ref{Fig:Phase}(c) and \ref{Fig:Phase}(d)]. These results highlight the nonuniform spin-orbit coupling generally existing in experiments but ignored in most simulations.

\begin{figure}[t!]
\includegraphics[width=0.85\columnwidth]{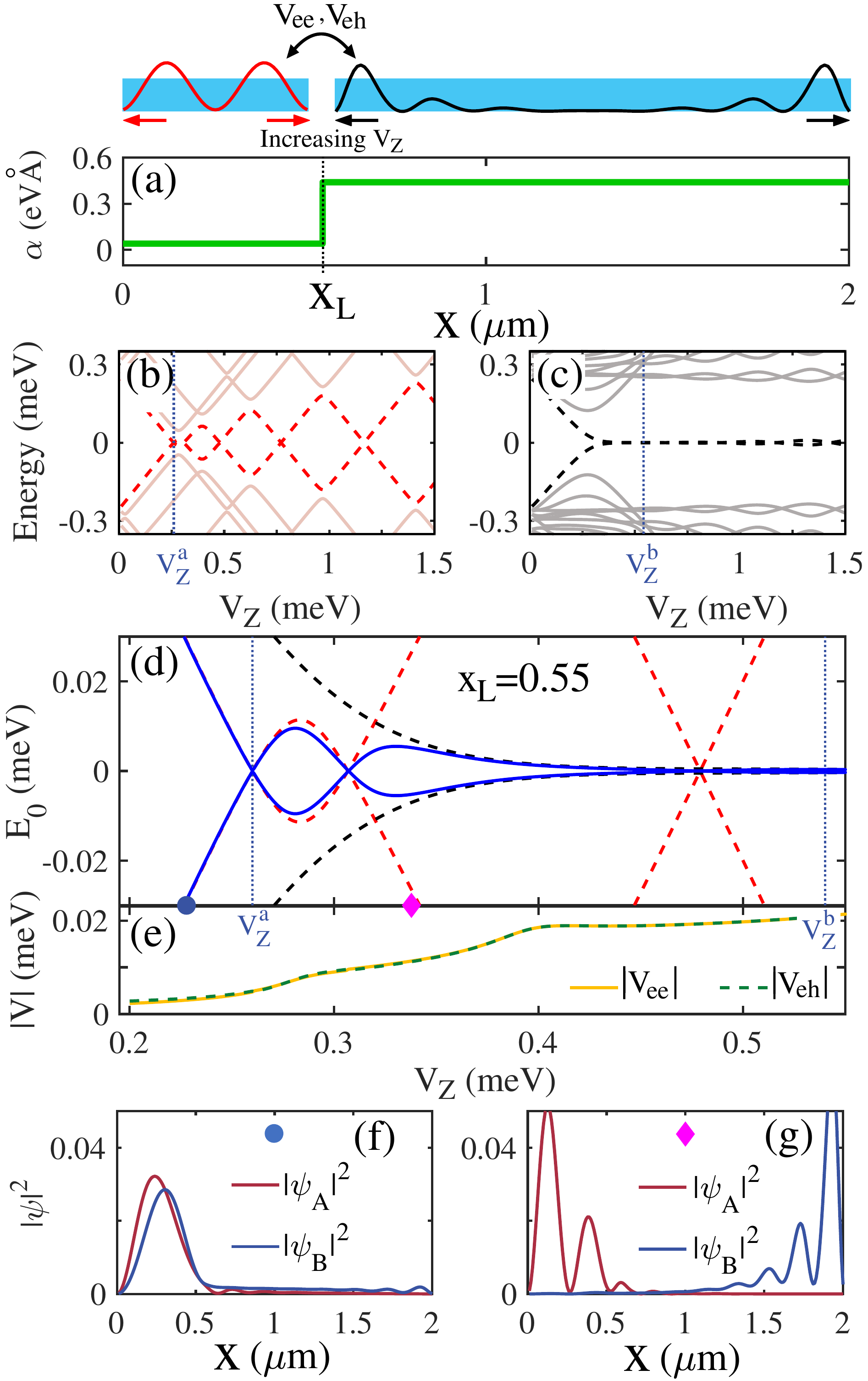}
\caption{Why Majorana oscillations decay. (a) A nanowire is decoupled at $x_L$ into two parts with different spin-orbit coupling described by $\alpha(x)$. [(b) and (c)] The energy spectra of the left and right parts, respectively. (d) The coupling $V_{ee/eh}$ between the lowest-energy spectra in (b) and (c) (red and black dashed) repels their lower energies to form the blue solid spectrum, qualitatively consistent with the oscillation pattern in Fig.~\ref{Fig:Decay}(c). (e) $V_{ee/eh}$ increase with increasing $V_Z=g_{\textrm{eff}}\mu_BB/2$ since the wave functions move towards the wire ends [see top of (a)], so they suppress the enhanced oscillations in (b) into decaying oscillations with increasing periods [blue solid in (d)]. [(f) and (g)] Majorana wave functions $\psi_{A}=(1/\sqrt{2})(\psi_{E_0}+\psi_{-E_0})$ and $\psi_{B}=(i/\sqrt{2})(\psi_{E_0}-\psi_{-E_0})$, with $\psi_{\pm E_0}$ the lowest-energy wave functions of the entire wire, at $V_Z$ indicated in (d). The parameters are $m^*=0.026m_e$, $\Delta=0.25$ meV, $\alpha_0=0.04$ eV{\AA}, $A=0.4$ eV{\AA}, and $\mu=0$.}\label{Fig:WhyDecay}
\end{figure}

{\color{blue}\emph{Why Majorana oscillations decay}.}-- Before showing the numerical simulations of the decaying Majorana oscillations in Fig.~\ref{Fig:Decay}, we first use Fig.~\ref{Fig:WhyDecay} to give the mechanism underneath. Suppose that a wire of $2$ $\mu$m is divided at $x_L=0.55$ $\mu$m into two uncoupled parts, with smaller ($L$) and larger ($R$) spin-orbit coupling, respectively [Fig.~\ref{Fig:WhyDecay}(a)]. Their energy spectra are quite different due to different length and spin-orbit coupling strength \cite{stanescu2013dimensional,mishmash2016approaching}: on the left [Fig.~\ref{Fig:WhyDecay}(b)], the enhanced oscillations emerge simultaneously after the first zero-energy crossing at $V_Z^a$; on the right [Fig.~\ref{Fig:WhyDecay}(c)], two near-zero-energy bound states develop after $V_Z^b$ ($>V_Z^a$).

Turning on the coupling between the two parts, the lowest-energy spectrum can be modeled by
\begin{equation}
H_{\textrm{eff}}=
\sum_{i=L,R}E_{i}c_{i}^{\dag}c_{i}+(V_{ee}c_{L}^{\dag}c_{R}+V_{eh}c_{L}^{\dag}c_{R}^{\dag}+H.c.),
\end{equation}
where $E_{L/R}$ stand for the lowest-energy spectra in Figs.~\ref{Fig:WhyDecay}(b) and \ref{Fig:WhyDecay}(c), and $V_{ee}$ and $V_{eh}$ are the particle-particle and particle-hole couplings (details in Sec.~SII of the Supplemental Material \cite{Supp}) between the lowest-energy states of the two parts [see top of Fig.~\ref{Fig:WhyDecay}(a)]. Between $V_Z^a$ and $V_Z^b$, the spectrum of the entire nanowire depends on the competition between $E_{L/R}$ and $V_{ee/eh}$. Figure \ref{Fig:WhyDecay}(b) shows that $E_L$ oscillates with increasing amplitude as a function of $V_Z$, consistent with the known result for uniform spin-orbit coupling \cite{sarma2012splitting}. Also, the Majorana wave functions are known to move towards the nanowire ends with increasing $V_Z$ \cite{huang2018etamorphosis} (Sec.~SII of the Supplemental Material \cite{Supp}), leading to stronger overlap between them. As a result, $V_{ee}$ and $V_{eh}$ increase with increasing $V_Z$ [Fig.~\ref{Fig:WhyDecay}(e)]. $V_{ee}$ and $V_{eh}$, as off-diagonal elements, can repel $\min\{E_L,E_R\}$ to lower energies. If the repulsion is strong enough to suppress the increasing amplitude of $\min\{E_L,E_R\}$, the lowest-energy spectrum of the entire nanowire will show the decaying oscillations [blue solid curves in Fig.~\ref{Fig:WhyDecay}(d)], {qualitatively consistent with the oscillation pattern shown in Fig.~\ref{Fig:Decay}(c). In contrast, there will be enhanced oscillations if the repulsion by $V_{ee/eh}$ is not strong enough. Therefore, the competition between $E_{L/R}$ and $V_{ee/eh}$ can account for the decaying or enhanced oscillations (Sec.~SIII of the Supplemental Material \cite{Supp}).

{\color{blue}\emph{Model}.}-- To verify our physical picture, we perform simulations by using the steplike spin-orbit coupling. We model the nanowire island by the Hamiltonian \cite{lutchyn2010majorana,oreg2010helical}
$H=\int_{0}^{L} dx\Psi^{\dag}(x){\cal H}\Psi(x)$,  ${\cal H}=\left[ p_{x}^{2}/2m^{\ast}-\mu(x)-\sigma_{y} \left\{  \alpha(x),p_{x}\right\}/2\hbar \right]  \tau_{z}+V_{Z}\sigma_{x}+\Delta\tau_{x}$, where $L$, $m^*$, $p_x=-i\hbar\partial_x$, $\Delta$, and $V_Z=g_{\textrm{eff}}\mu_B B/2$ are the wire length, effective electron mass, momentum operator, effective pairing, and Zeeman energy induced by $B$, respectively. $g_{\textrm{eff}}$ and $\mu_B$ are the effective $g$ factor and Bohr magneton. $\mu(x)$ and $\alpha(x)$ denote the position-dependent chemical potential and spin-orbit coupling, respectively. Quite different from the previous theories which assume a constant spin-orbit coupling \cite{chiu2017conductance,dmytruk2018suppression,fleckenstein2018decaying}, we model that spin-orbit coupling has a profile [see also the green curve in Fig.~\ref{Fig:Decay}(a)]
\begin{equation}
\alpha(x)=\frac{A}{2}\left[\tanh\left(\frac{x-x_L}{\lambda_L}\right)+\tanh\left(\frac{x_R-x}{\lambda_R}\right)\right]+\alpha_0, \label{eq1}
\end{equation}
where $A$, $\alpha_0$, $x_{L/R}$, and $\lambda_{L/R}$ are the parameters that describe the profile. ${\cal H}$ is written in terms of the Nambu spinor $\{u_\uparrow(x),u_\downarrow(x),v_\downarrow(x),-v_\uparrow(x)\}$. The Pauli matrices $\sigma$ and $\tau$ act on the spin and particle-hole spaces, respectively. The anticommutator in ${\cal H}$ ensures the Hermiticity \cite{sanchez2006fano,klinovaja2015fermionic,dolcini2018magnetic}. In realistic experiments, the parameters intertwine when changing the gate voltages \cite{vuik2016effects,bommer2018spin,woods2018effective,mikkelsen2018hybridization,antipov2018effects,vaitiekenas2018effective,de2018electric}, and the superconductor can induce renormalization effects \cite{stanescu2011majorana,reeg2018metallization}. Nevertheless, to focus on the effect of the steplike spin-orbit coupling, all the parameters in $H$ are assumed to be independently adjustable. By diagonalizing $H$ on a lattice, the energy spectrum and wave functions are obtained. The lowest energy is the bound state energy $E_0$, the hybridization energy mentioned above.

{\color{blue}\emph{Decays of Majorana oscillations}.}-- The blue curves in Figs.~\ref{Fig:Decay}(b)-(d) show our numerical results using three sets of model parameters listed in Sec.~SI of the Supplemental Material \cite{Supp}. To focus on the effect of the steplike spin-orbit coupling, first we consider only one step of spin-orbit coupling, so that $\alpha(x)=\alpha_0+A\Theta(x-x_L)$ [Fig.~\ref{Fig:WhyDecay}(a)]; i.e., let $x_R=L$ and $\lambda_L=\lambda_R=a$ in Eq.~(\ref{eq1}). Our simulations agree with the experiments, not only for the decaying amplitude, but also including the lowest-energy crossing [Figs.~\ref{Fig:Decay}(b) and \ref{Fig:Decay}(c)], anticrossing [Fig.~\ref{Fig:Decay}(d)], and increasing oscillation period in a magnetic field [Fig.~\ref{Fig:Decay}(c)]. We note that our results are generic and do not depend on the detailed parameters, e.g., the step shape (smoothness), effective pairing $\Delta$, chemical potential $\mu$, and spin-orbit coupling strength (Secs.~SIII and SIV of the Supplemental Material \cite{Supp}). Further increasing the magnetic field, the oscillations may turn from decay to increase for those magnetic fields at which the superconductivity is suppressed in the experiments, thus less likely to be observed (see Figs.~S2 and S3 in the Supplemental Material \cite{Supp}).

{\color{blue}\emph{Decays of Andreev oscillations}.}-- Are these decaying oscillations unique for Majorana bound states? Our answer is no. It has been suggested that the same device can also host the Andreev bound states \cite{kells2012near,stanescu2013disentangling,liu2017andreev,moore2018two,vuik2018reproducing}. Whether the decaying oscillations are from Andreev or Majorana bound states can be checked from the spatial profiles of the lowest-energy Majorana wave functions at the Zeeman energies indicated in Fig.~\ref{Fig:WhyDecay}(d). The Majorana wave functions can be constructed by projecting the lowest-energy wave functions onto the Majorana basis \cite{kjaergaard2012majorana,moore2018two}, i.e., $\psi_{A}=(1/\sqrt{2})(\psi_{E_0}+\psi_{-E_0})$ and $\psi_{B}=(i/\sqrt{2})(\psi_{E_0}-\psi_{-E_0})$. $\psi_A$ and $\psi_B$ are localized at the opposite wire ends for the Majorana bound states, while they are strongly overlapping or separated by a distance comparable with the penetration length for the Andreev bound states \cite{moore2018two}. For $V_Z$ far smaller than $V_Z^a$ [Fig.~\ref{Fig:WhyDecay}(f)], the two wave functions are squeezed in the region with small spin-orbit coupling ($0<x<x_L$), implying that they are two Andreev bound states. For $V_Z$ larger than $V_Z^a$ [Fig.~\ref{Fig:WhyDecay}(g)], the two wave functions are well localized at the opposite ends, forming a pair of near-zero-energy Majorana bound states with a slight overlap.

\begin{figure}[t!]
\includegraphics[width=0.9\columnwidth]{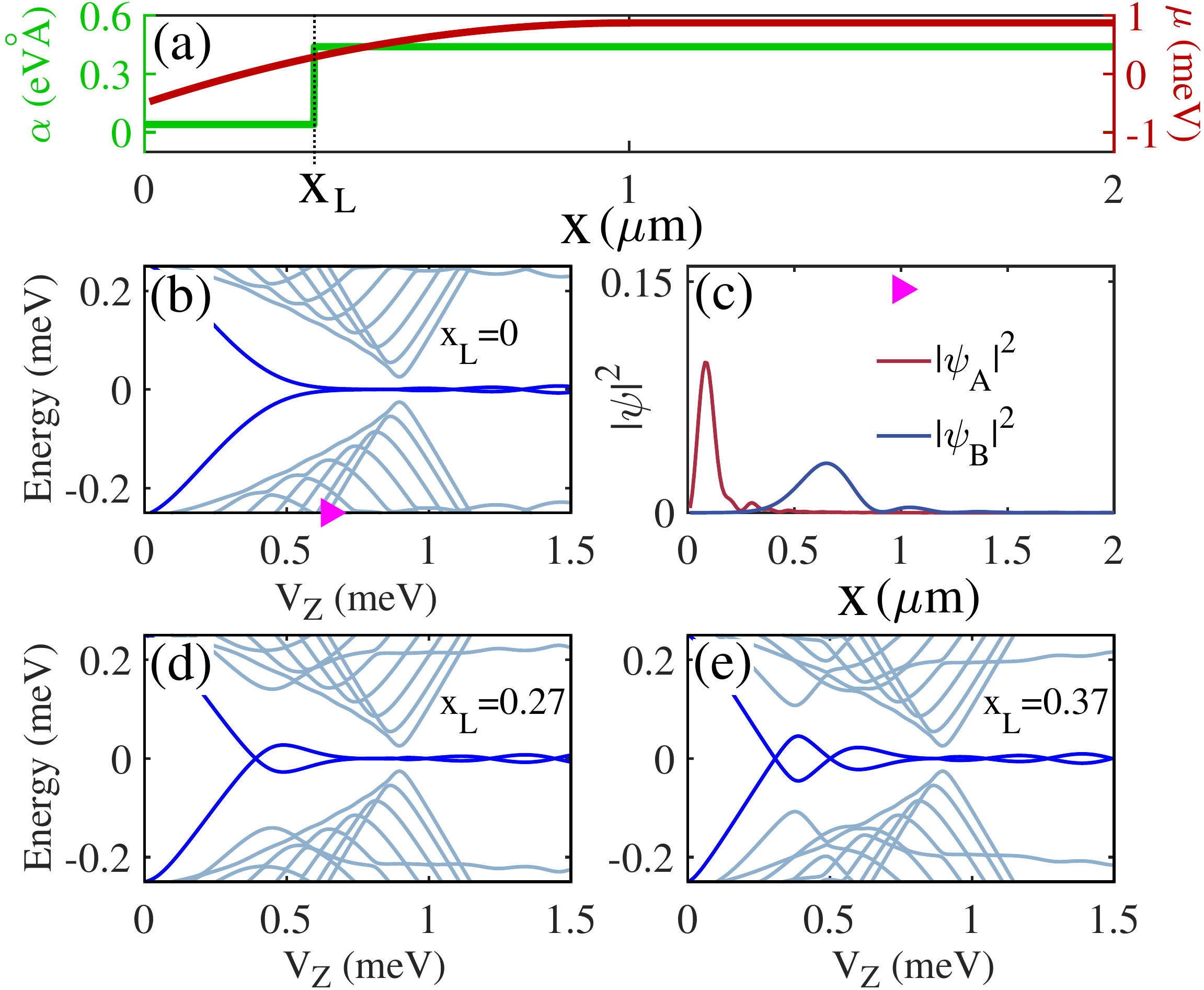}
\caption{(a) The steplike spin-orbit coupling $\alpha(x)$ and smoothly varying chemical potential $\mu(x)$. (b) $x_L=0$ means a uniform spin-orbit coupling, for which, the near-zero-energy Andreev bound states persist for a wide range of $V_Z$ before the topological phase transition point at which the superconducting gap nearly closes and reopens. (c) At $V_Z$ marked by the triangle in (b), the projections of the lowest-energy wave functions on the Majorana basis are partially separated. [(d) and (e)] In the presence of the steplike spin-orbit coupling with different $x_L$, decaying oscillations also exist in the spectrum of the Andreev bound states (trivial regime). The parameters except for $\mu(x)$ are the same as those in Fig.~\ref{Fig:WhyDecay}.}\label{Fig:Andreev}
\end{figure}

We simulate the near-zero-energy Andreev bound states by employing a smoothly varying chemical potential $\mu(x)$ \cite{kells2012near,stanescu2013disentangling,liu2017andreev,moore2018two,vuik2018reproducing}, as shown in Fig.~\ref{Fig:Andreev}(a). For uniform spin-orbit coupling (i.e., $x_L=0$), two near-zero-energy bound states persist over a wide range of Zeeman energy before the topological phase transition point $V_Z^C=\sqrt{\textrm{max}|\mu(x)|^2+\Delta^2}$ [about 0.91 meV in Fig.~\ref{Fig:Andreev}(b)] at which the superconducting gap nearly closes and reopens. These bound states are partially separated Andreev bound states \cite{moore2018two} since the constituent Majorana wave functions are separated by a distance comparable with the penetration length [Fig.~\ref{Fig:Andreev}(c)]. After including a steplike distribution of spin-orbit coupling, Figs.~\ref{Fig:Andreev}(d) and \ref{Fig:Andreev}(e) show that there are also decaying oscillations for the Andreev bound states at $V_Z<V_Z^C$. The oscillations turn to increase at $V_Z>V_Z^C$ for Majorana bound states. The Andreev or Majorana nature is determined by the spatial profiles of the projections of the lowest-energy wave functions onto the Majorana basis and these decaying oscillations are also due to the competition between $E_{L/R}$ and $V_{ee/eh}$, similar to Fig.~\ref{Fig:WhyDecay}(d) (Sec.~SV of the Supplemental Material \cite{Supp}).

\begin{figure}[t!]
\includegraphics[width=0.9\columnwidth]{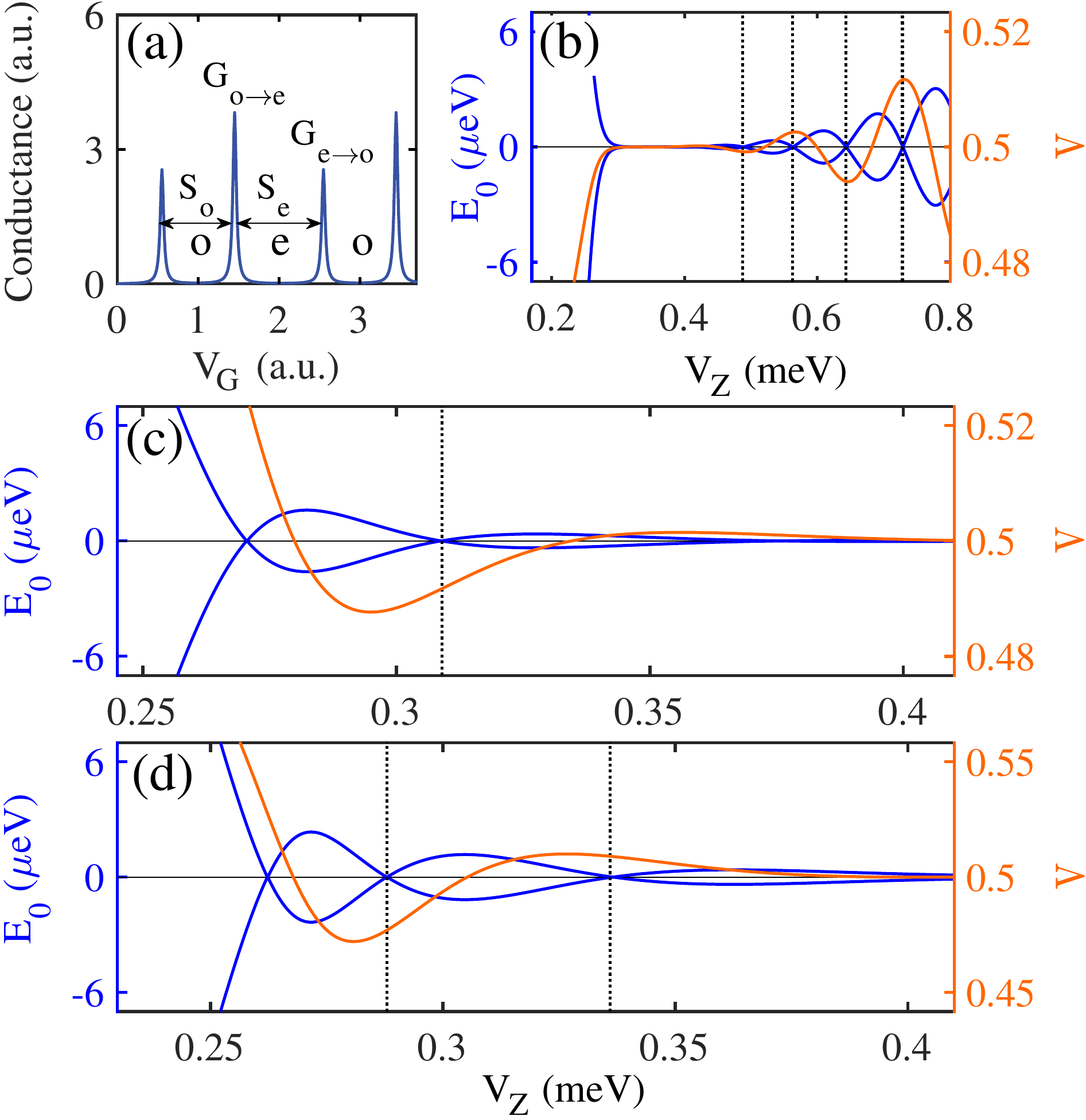}
\caption{(a) The Coulomb blockade peaks of the conductance as a function of the gate voltage $V_{\mathrm{G}}$. The transition from the odd to even parity ($o\rightarrow e$) differs from $e\rightarrow o$ in the height of peak ($G_{o\rightarrow e}$, $G_{e\rightarrow o}$) and spacing between two neighboring peaks ($S_e$, $S_o$). The experimental measured $\pm E_0$ are extracted from the peak spacings $\pm E_0=\eta \langle S_{e(o)} \rangle-E_C$ [corresponding to the red and blue data in Figs.~\ref{Fig:Decay}(b)-\ref{Fig:Decay}(d)], where $\langle S_e\rangle$ and $\langle S_o\rangle$ are the ensemble-averaged peak spacings, $E_C$ the charging energy, and $\eta=2E_C/(\langle S_e\rangle+\langle S_o\rangle)$ (Sec.~SVI of the Supplemental Material \cite{Supp}). Lowest-energy spectrum $E_0$ (blue curves) and peak height ratio $\Lambda=G_{e\rightarrow o}/(G_{e\rightarrow o}+G_{o\rightarrow e})$ (orange curves) as functions of $V_Z$ with (b) constant spin-orbit coupling $\alpha=0.16$ eV{\AA} and (c),(d) smoothed steplike spin-orbit coupling described by Eq.~(\ref{eq1}) with (c) $x_L=0.43$ $\mu$m, $x_R=2$ $\mu$m, $\alpha_0=0.03$ eV{\AA}, and $A=0.44$ eV{\AA}, and (d) $x_L=0.34$ $\mu$m, $x_R=1.8$ $\mu$m, $\alpha_0=0.02$ eV{\AA}, and $A=0.5$ eV{\AA}. Other parameters are $L=2.3$ $\mu$m, $\lambda_L=\lambda_R=0.05$ $\mu$m, $\mu=0$, and $\Gamma_L=\Gamma_R$. Here the steps of spin-orbit coupling are smoothed by using finite $\lambda_{L/R}$. }\label{Fig:Phase}
\end{figure}

{\color{blue}\emph{Phase shift between peak spacing and height oscillations}.}-- In the floating nanowire island [Fig.~\ref{Fig:Decay}(a)] \cite{fu2010electron,hutzen2012majorana,higginbotham2015parity,lu2016enhanced,van2016conductance}, adding an electron costs a finite charging energy due to its small capacitance \cite{grabert2013single}, leading to the Coulomb blockade peaks in the two-terminal conductance measurement [Fig.~\ref{Fig:Phase}(a)]. Because of the hybridization energy $E_0$, charging a pair of unoccupied Majorana bound states to occupied ($e\rightarrow o$) differs in energy from the process $o\rightarrow e$ in the next charging event. In this way, $E_0$ can be extracted from the difference between two consecutive Coulomb blockade peak spacings in gate voltage (Sec.~SVI of the Supplemental Material \cite{Supp}). The blue curves in Figs.~\ref{Fig:Phase}(c) and \ref{Fig:Phase}(d) show the calculated decaying Majorana oscillations of $E_0$. Different from Fig.~\ref{Fig:WhyDecay}, here we consider two steps of spin-orbit coupling and the steps are smoothed by using finite $\lambda_{L/R}$, as depicted in Fig.~\ref{Fig:Decay}(a). In addition, Figs.~\ref{Fig:Phase}(b)-\ref{Fig:Phase}(d) also present the calculated Coulomb blockade peak height ratio $\Lambda=G_{e\rightarrow o}/(G_{e\rightarrow o}+G_{o\rightarrow e})$ as a function of the Zeeman energy $V_Z$ (orange curves). The corresponding conductance peak heights $G_{e\rightarrow o}$ and $G_{o\rightarrow e}$ are shown in Sec.~SVI of the Supplemental Material \cite{Supp}. The zero-temperature peak heights are assumed independent of $V_G$ and are formulated as $G_{e\rightarrow o}$ = $(e^2/\hbar)(\Gamma_L\Gamma_R|u_L|^2|u_R|^2)$/$(\Gamma_L|u_L|^2+\Gamma_R|u_R|^2)$ \cite{hansen2018probing}, where $\Gamma_{L(R)}$ is the tunneling rate between the left (right) end of the nanowire and its nearest metallic lead, and $|u_{L(R)}|^2=\sum_{\sigma=\uparrow,\downarrow}|u_{L(R)\sigma}|^2$ with $u_{L(R)\sigma}$ the lowest-energy wave function component at the leftmost (rightmost) lattice site of the wire. $G_{o\rightarrow e}$ is obtained by replacing all $u_{L(R)\sigma}$ in $G_{e\rightarrow o}$ with $v_{L(R)\sigma}$, which means that $G_{e\rightarrow o}$ and $G_{o\rightarrow e}$ are related to the electronlike and holelike components of the lowest-energy state, respectively. It has been predicted \cite{hansen2018probing} that the oscillations of $\Lambda$ are correlated to those of $E_0$ by a $\pi/2$ phase shift for the Majorana bound states. Specifically, $E_0$ is zero at the extremals of $\Lambda$, and $\Lambda=1/2$ at the extremals of $E_0$ [Fig.~\ref{Fig:Phase}(b)]. While for the Andreev bound states, there is no such correlated $\pi/2$ phase shift \cite{hansen2018probing}. When considering the steplike spin-orbit coupling in our model, the correlations for our decaying Majorana oscillations show clear deviations from the exact $\pi/2$ phase shift [Figs.~\ref{Fig:Phase}(c) and \ref{Fig:Phase}(d)]. This implies that the steplike spin-orbit coupling may be one of the reasons why the correlation between $E_0$ and $\Lambda$ is ambiguous in a recent experiment \cite{shen2018parity}, since not only the Andreev bound states, but also the Majorana states can give uncorrelated oscillation patterns between $E_0$ and $\Lambda$ when spin-orbit coupling is nonuniform. Nonuniform spin-orbit coupling has recently been studied in a different context \cite{Reeg18zero}, in which a spin-orbit-coupled quantum dot is attached to a zero spin-orbit coupling nanowire, leading to localized zero-energy Andreev bound states in the quantum dot.

We thank helpful discussions with Wen-Yu Shan. This work was supported by the Strategic Priority Research Program of Chinese Academy of Sciences (Grant No. XDB28000000), the Guangdong Innovative and Entrepreneurial Research Team Program (2016ZT06D348), the National Basic Research Program of China (2015CB921102), the National Key R \& D Program (2016YFA0301700), the National Natural Science Foundation of China (11534001, 11574127, 61474018), and the Science, Technology and Innovation Commission of Shenzhen Municipality (ZDSYS20170303165926217,  JCYJ20170412152620376).


\begin{thebibliography}{73}%
\makeatletter
\providecommand \@ifxundefined [1]{%
 \@ifx{#1\undefined}
}%
\providecommand \@ifnum [1]{%
 \ifnum #1\expandafter \@firstoftwo
 \else \expandafter \@secondoftwo
 \fi
}%
\providecommand \@ifx [1]{%
 \ifx #1\expandafter \@firstoftwo
 \else \expandafter \@secondoftwo
 \fi
}%
\providecommand \natexlab [1]{#1}%
\providecommand \enquote  [1]{``#1''}%
\providecommand \bibnamefont  [1]{#1}%
\providecommand \bibfnamefont [1]{#1}%
\providecommand \citenamefont [1]{#1}%
\providecommand \href@noop [0]{\@secondoftwo}%
\providecommand \href [0]{\begingroup \@sanitize@url \@href}%
\providecommand \@href[1]{\@@startlink{#1}\@@href}%
\providecommand \@@href[1]{\endgroup#1\@@endlink}%
\providecommand \@sanitize@url [0]{\catcode `\\12\catcode `\$12\catcode
  `\&12\catcode `\#12\catcode `\^12\catcode `\_12\catcode `\%12\relax}%
\providecommand \@@startlink[1]{}%
\providecommand \@@endlink[0]{}%
\providecommand \url  [0]{\begingroup\@sanitize@url \@url }%
\providecommand \@url [1]{\endgroup\@href {#1}{\urlprefix }}%
\providecommand \urlprefix  [0]{URL }%
\providecommand \Eprint [0]{\href }%
\providecommand \doibase [0]{http://dx.doi.org/}%
\providecommand \selectlanguage [0]{\@gobble}%
\providecommand \bibinfo  [0]{\@secondoftwo}%
\providecommand \bibfield  [0]{\@secondoftwo}%
\providecommand \translation [1]{[#1]}%
\providecommand \BibitemOpen [0]{}%
\providecommand \bibitemStop [0]{}%
\providecommand \bibitemNoStop [0]{.\EOS\space}%
\providecommand \EOS [0]{\spacefactor3000\relax}%
\providecommand \BibitemShut  [1]{\csname bibitem#1\endcsname}%
\let\auto@bib@innerbib\@empty
\bibitem [{\citenamefont {Alicea}(2012)}]{alicea2012new}%
  \BibitemOpen
  \bibfield  {author} {\bibinfo {author} {\bibfnamefont {J.}~\bibnamefont
  {Alicea}},\ }\bibfield  {title} {\enquote {\bibinfo {title} {New directions
  in the pursuit of {Majorana} fermions in solid state systems}}, }\href
  {https://doi.org/10.1088/0034-4885/75/7/076501} {\bibfield  {journal}
  {\bibinfo  {journal} {Rep. Prog. Phys.}\ }\textbf {\bibinfo {volume} {75}},\
  \bibinfo {pages} {076501} (\bibinfo {year} {2012})}\BibitemShut {NoStop}%
\bibitem [{\citenamefont {Leijnse}\ and\ \citenamefont
  {Flensberg}(2012)}]{leijnse2012introduction}%
  \BibitemOpen
  \bibfield  {author} {\bibinfo {author} {\bibfnamefont {M.}~\bibnamefont
  {Leijnse}}\ and\ \bibinfo {author} {\bibfnamefont {K.}~\bibnamefont
  {Flensberg}},\ }\bibfield  {title} {\enquote {\bibinfo {title} {Introduction
  to topological superconductivity and {Majorana} fermions}}, }\href
  {https://doi.org/10.1088/0268-1242/27/12/124003} {\bibfield  {journal}
  {\bibinfo  {journal} {Semicond. Sci. Technol.}\ }\textbf {\bibinfo {volume}
  {27}},\ \bibinfo {pages} {124003} (\bibinfo {year} {2012})}\BibitemShut
  {NoStop}%
\bibitem [{\citenamefont {Beenakker}(2013)}]{beenakker2013search}%
  \BibitemOpen
  \bibfield  {author} {\bibinfo {author} {\bibfnamefont {C.}~\bibnamefont
  {Beenakker}},\ }\bibfield  {title} {\enquote {\bibinfo {title} {Search for
  {Majorana} fermions in superconductors}}, }\href
  {https://doi.org/10.1146/annurev-conmatphys-030212-184337} {\bibfield
  {journal} {\bibinfo  {journal} {Annu. Rev. Condens. Matter Phys.}\ }\textbf
  {\bibinfo {volume} {4}},\ \bibinfo {pages} {113} (\bibinfo {year}
  {2013})}\BibitemShut {NoStop}%
\bibitem [{\citenamefont {Stanescu}\ and\ \citenamefont
  {Tewari}(2013{\natexlab{a}})}]{stanescu2013majorana}%
  \BibitemOpen
  \bibfield  {author} {\bibinfo {author} {\bibfnamefont {T.~D.}\ \bibnamefont
  {Stanescu}}\ and\ \bibinfo {author} {\bibfnamefont {S.}~\bibnamefont
  {Tewari}},\ }\bibfield  {title} {\enquote {\bibinfo {title} {{Majorana}
  fermions in semiconductor nanowires: fundamentals, modeling, and
  experiment}}, }\href {https://doi.org/10.1088/0953-8984/25/23/233201}
  {\bibfield  {journal} {\bibinfo  {journal} {J. Phys.: Condens. Matter}\
  }\textbf {\bibinfo {volume} {25}},\ \bibinfo {pages} {233201} (\bibinfo
  {year} {2013}{\natexlab{a}})}\BibitemShut {NoStop}%
\bibitem [{\citenamefont {Aguado}(2017)}]{aguado2017majorana}%
  \BibitemOpen
  \bibfield  {author} {\bibinfo {author} {\bibfnamefont {R.}~\bibnamefont
  {Aguado}},\ }\bibfield  {title} {\enquote {\bibinfo {title} {{Majorana}
  quasiparticles in condensed matter}}, }\href
  {https://doi.org/10.1393/ncr/i2017-10141-9} {\bibfield  {journal} {\bibinfo
  {journal} {Riv. Nuovo Cimento}\ }\textbf {\bibinfo {volume} {40}},\ \bibinfo
  {pages} {523} (\bibinfo {year} {2017})}\BibitemShut {NoStop}%
\bibitem [{\citenamefont {Kitaev}(2003)}]{kitaev2003fault}%
  \BibitemOpen
  \bibfield  {author} {\bibinfo {author} {\bibfnamefont {A.~Y.}\ \bibnamefont
  {Kitaev}},\ }\bibfield  {title} {\enquote {\bibinfo {title} {Fault-tolerant
  quantum computation by anyons}}, }\href
  {https://doi.org/10.1016/S0003-4916(02)00018-0} {\bibfield  {journal}
  {\bibinfo  {journal} {Ann. Phys. (Amsterdam)}\ }\textbf {\bibinfo {volume}
  {303}},\ \bibinfo {pages} {2} (\bibinfo {year} {2003})}\BibitemShut {NoStop}%
\bibitem [{\citenamefont {Nayak}\ \emph {et~al.}(2008)\citenamefont {Nayak},
  \citenamefont {Simon}, \citenamefont {Stern}, \citenamefont {Freedman},\ and\
  \citenamefont {{Das Sarma}}}]{nayak2008non}%
  \BibitemOpen
  \bibfield  {author} {\bibinfo {author} {\bibfnamefont {C.}~\bibnamefont
  {Nayak}}, \bibinfo {author} {\bibfnamefont {S.~H.}\ \bibnamefont {Simon}},
  \bibinfo {author} {\bibfnamefont {A.}~\bibnamefont {Stern}}, \bibinfo
  {author} {\bibfnamefont {M.}~\bibnamefont {Freedman}}, \ and\ \bibinfo
  {author} {\bibfnamefont {S.}~\bibnamefont {{Das Sarma}}},\ }\bibfield
  {title} {\enquote {\bibinfo {title} {Non-{Abelian} anyons and topological
  quantum computation}}, }\href {https://doi.org/10.1103/RevModPhys.80.1083}
  {\bibfield  {journal} {\bibinfo  {journal} {Rev. Mod. Phys.}\ }\textbf
  {\bibinfo {volume} {80}},\ \bibinfo {pages} {1083} (\bibinfo {year}
  {2008})}\BibitemShut {NoStop}%
\bibitem [{\citenamefont {{Das Sarma}}\ \emph {et~al.}(2015)\citenamefont {{Das
  Sarma}}, \citenamefont {Freedman},\ and\ \citenamefont
  {Nayak}}]{sarma2015majorana}%
  \BibitemOpen
  \bibfield  {author} {\bibinfo {author} {\bibfnamefont {S.}~\bibnamefont {{Das
  Sarma}}}, \bibinfo {author} {\bibfnamefont {M.}~\bibnamefont {Freedman}}, \
  and\ \bibinfo {author} {\bibfnamefont {C.}~\bibnamefont {Nayak}},\ }\bibfield
   {title} {\enquote {\bibinfo {title} {{Majorana} zero modes and topological
  quantum computation}}, }\href {https://doi.org/10.1038/npjqi.2015.1}
  {\bibfield  {journal} {\bibinfo  {journal} {npj Quantum Inf.}\ }\textbf
  {\bibinfo {volume} {1}},\ \bibinfo {pages} {15001} (\bibinfo {year}
  {2015})}\BibitemShut {NoStop}%
\bibitem [{\citenamefont {Lutchyn}\ \emph {et~al.}(2010)\citenamefont
  {Lutchyn}, \citenamefont {Sau},\ and\ \citenamefont {{Das
  Sarma}}}]{lutchyn2010majorana}%
  \BibitemOpen
  \bibfield  {author} {\bibinfo {author} {\bibfnamefont {R.~M.}\ \bibnamefont
  {Lutchyn}}, \bibinfo {author} {\bibfnamefont {J.~D.}\ \bibnamefont {Sau}}, \
  and\ \bibinfo {author} {\bibfnamefont {S.}~\bibnamefont {{Das Sarma}}},\
  }\bibfield  {title} {\enquote {\bibinfo {title} {{Majorana} fermions and a
  topological phase transition in semiconductor-superconductor
  heterostructures}}, }\href {https://doi.org/10.1103/PhysRevLett.105.077001}
  {\bibfield  {journal} {\bibinfo  {journal} {Phys. Rev. Lett.}\ }\textbf
  {\bibinfo {volume} {105}},\ \bibinfo {pages} {077001} (\bibinfo {year}
  {2010})}\BibitemShut {NoStop}%
\bibitem [{\citenamefont {Oreg}\ \emph {et~al.}(2010)\citenamefont {Oreg},
  \citenamefont {Refael},\ and\ \citenamefont {von Oppen}}]{oreg2010helical}%
  \BibitemOpen
  \bibfield  {author} {\bibinfo {author} {\bibfnamefont {Y.}~\bibnamefont
  {Oreg}}, \bibinfo {author} {\bibfnamefont {G.}~\bibnamefont {Refael}}, \ and\
  \bibinfo {author} {\bibfnamefont {F.}~\bibnamefont {von Oppen}},\ }\bibfield
  {title} {\enquote {\bibinfo {title} {Helical liquids and {Majorana} bound
  states in quantum wires}}, }\href
  {https://doi.org/10.1103/PhysRevLett.105.177002} {\bibfield  {journal}
  {\bibinfo  {journal} {Phys. Rev. Lett.}\ }\textbf {\bibinfo {volume} {105}},\
  \bibinfo {pages} {177002} (\bibinfo {year} {2010})}\BibitemShut {NoStop}%
\bibitem [{\citenamefont {Mourik}\ \emph {et~al.}(2012)\citenamefont {Mourik},
  \citenamefont {Zuo}, \citenamefont {Frolov}, \citenamefont {Plissard},
  \citenamefont {Bakkers},\ and\ \citenamefont
  {Kouwenhoven}}]{mourik2012signatures}%
  \BibitemOpen
  \bibfield  {author} {\bibinfo {author} {\bibfnamefont {V.}~\bibnamefont
  {Mourik}}, \bibinfo {author} {\bibfnamefont {K.}~\bibnamefont {Zuo}},
  \bibinfo {author} {\bibfnamefont {S.~M.}\ \bibnamefont {Frolov}}, \bibinfo
  {author} {\bibfnamefont {S.}~\bibnamefont {Plissard}}, \bibinfo {author}
  {\bibfnamefont {E.~P.}\ \bibnamefont {Bakkers}}, \ and\ \bibinfo {author}
  {\bibfnamefont {L.~P.}\ \bibnamefont {Kouwenhoven}},\ }\bibfield  {title}
  {\enquote {\bibinfo {title} {Signatures of {Majorana} fermions in hybrid
  superconductor-semiconductor nanowire devices}}, }\href
  {https://doi.org/10.1126/science.1222360} {\bibfield  {journal} {\bibinfo
  {journal} {Science}\ }\textbf {\bibinfo {volume} {336}},\ \bibinfo {pages}
  {1003} (\bibinfo {year} {2012})}\BibitemShut {NoStop}%
\bibitem [{\citenamefont {Deng}\ \emph {et~al.}(2012)\citenamefont {Deng},
  \citenamefont {Yu}, \citenamefont {Huang}, \citenamefont {Larsson},
  \citenamefont {Caroff},\ and\ \citenamefont {Xu}}]{deng2012anomalous}%
  \BibitemOpen
  \bibfield  {author} {\bibinfo {author} {\bibfnamefont {M.}~\bibnamefont
  {Deng}}, \bibinfo {author} {\bibfnamefont {C.}~\bibnamefont {Yu}}, \bibinfo
  {author} {\bibfnamefont {G.}~\bibnamefont {Huang}}, \bibinfo {author}
  {\bibfnamefont {M.}~\bibnamefont {Larsson}}, \bibinfo {author} {\bibfnamefont
  {P.}~\bibnamefont {Caroff}}, \ and\ \bibinfo {author} {\bibfnamefont
  {H.}~\bibnamefont {Xu}},\ }\bibfield  {title} {\enquote {\bibinfo {title}
  {Anomalous zero-bias conductance peak in a {Nb--InSb nanowire--Nb} hybrid
  device}}, }\href {https://doi.org/10.1021/nl303758w} {\bibfield  {journal}
  {\bibinfo  {journal} {Nano Lett.}\ }\textbf {\bibinfo {volume} {12}},\
  \bibinfo {pages} {6414} (\bibinfo {year} {2012})}\BibitemShut {NoStop}%
\bibitem [{\citenamefont {Das}\ \emph {et~al.}(2012)\citenamefont {Das},
  \citenamefont {Ronen}, \citenamefont {Most}, \citenamefont {Oreg},
  \citenamefont {Heiblum},\ and\ \citenamefont {Shtrikman}}]{das2012zero}%
  \BibitemOpen
  \bibfield  {author} {\bibinfo {author} {\bibfnamefont {A.}~\bibnamefont
  {Das}}, \bibinfo {author} {\bibfnamefont {Y.}~\bibnamefont {Ronen}}, \bibinfo
  {author} {\bibfnamefont {Y.}~\bibnamefont {Most}}, \bibinfo {author}
  {\bibfnamefont {Y.}~\bibnamefont {Oreg}}, \bibinfo {author} {\bibfnamefont
  {M.}~\bibnamefont {Heiblum}}, \ and\ \bibinfo {author} {\bibfnamefont
  {H.}~\bibnamefont {Shtrikman}},\ }\bibfield  {title} {\enquote {\bibinfo
  {title} {Zero-bias peaks and splitting in an {Al--InAs} nanowire topological
  superconductor as a signature of {Majorana} fermions}}, }\href
  {https://doi.org/10.1038/nphys2479} {\bibfield  {journal} {\bibinfo
  {journal} {Nat. Phys.}\ }\textbf {\bibinfo {volume} {8}},\ \bibinfo {pages}
  {887} (\bibinfo {year} {2012})}\BibitemShut {NoStop}%
\bibitem [{\citenamefont {Finck}\ \emph {et~al.}(2013)\citenamefont {Finck},
  \citenamefont {Van~Harlingen}, \citenamefont {Mohseni}, \citenamefont
  {Jung},\ and\ \citenamefont {Li}}]{finck2013anomalous}%
  \BibitemOpen
  \bibfield  {author} {\bibinfo {author} {\bibfnamefont {A.~D.~K.}\
  \bibnamefont {Finck}}, \bibinfo {author} {\bibfnamefont {D.~J.}\ \bibnamefont
  {Van~Harlingen}}, \bibinfo {author} {\bibfnamefont {P.~K.}\ \bibnamefont
  {Mohseni}}, \bibinfo {author} {\bibfnamefont {K.}~\bibnamefont {Jung}}, \
  and\ \bibinfo {author} {\bibfnamefont {X.}~\bibnamefont {Li}},\ }\bibfield
  {title} {\enquote {\bibinfo {title} {Anomalous modulation of a zero-bias peak
  in a hybrid nanowire-superconductor device}}, }\href {\doibase
  10.1103/PhysRevLett.110.126406} {\bibfield  {journal} {\bibinfo  {journal}
  {Phys. Rev. Lett.}\ }\textbf {\bibinfo {volume} {110}},\ \bibinfo {pages}
  {126406} (\bibinfo {year} {2013})}\BibitemShut {NoStop}%
\bibitem [{\citenamefont {Churchill}\ \emph {et~al.}(2013)\citenamefont
  {Churchill}, \citenamefont {Fatemi}, \citenamefont {Grove-Rasmussen},
  \citenamefont {Deng}, \citenamefont {Caroff}, \citenamefont {Xu},\ and\
  \citenamefont {Marcus}}]{churchill2013superconductor}%
  \BibitemOpen
  \bibfield  {author} {\bibinfo {author} {\bibfnamefont {H.~O.~H.}\
  \bibnamefont {Churchill}}, \bibinfo {author} {\bibfnamefont {V.}~\bibnamefont
  {Fatemi}}, \bibinfo {author} {\bibfnamefont {K.}~\bibnamefont
  {Grove-Rasmussen}}, \bibinfo {author} {\bibfnamefont {M.~T.}\ \bibnamefont
  {Deng}}, \bibinfo {author} {\bibfnamefont {P.}~\bibnamefont {Caroff}},
  \bibinfo {author} {\bibfnamefont {H.~Q.}\ \bibnamefont {Xu}}, \ and\ \bibinfo
  {author} {\bibfnamefont {C.~M.}\ \bibnamefont {Marcus}},\ }\bibfield  {title}
  {\enquote {\bibinfo {title} {Superconductor-nanowire devices from tunneling
  to the multichannel regime: Zero-bias oscillations and magnetoconductance
  crossover}}, }\href {\doibase 10.1103/PhysRevB.87.241401} {\bibfield
  {journal} {\bibinfo  {journal} {Phys. Rev. B}\ }\textbf {\bibinfo {volume}
  {87}},\ \bibinfo {pages} {241401(R)} (\bibinfo {year} {2013})}\BibitemShut
  {NoStop}%
\bibitem [{\citenamefont {Deng}\ \emph {et~al.}(2016)\citenamefont {Deng},
  \citenamefont {Vaitiek{\.e}nas}, \citenamefont {Hansen}, \citenamefont
  {Danon}, \citenamefont {Leijnse}, \citenamefont {Flensberg}, \citenamefont
  {Nyg{\aa}rd}, \citenamefont {Krogstrup},\ and\ \citenamefont
  {Marcus}}]{deng2016majorana}%
  \BibitemOpen
  \bibfield  {author} {\bibinfo {author} {\bibfnamefont {M.}~\bibnamefont
  {Deng}}, \bibinfo {author} {\bibfnamefont {S.}~\bibnamefont
  {Vaitiek{\.e}nas}}, \bibinfo {author} {\bibfnamefont {E.~B.}\ \bibnamefont
  {Hansen}}, \bibinfo {author} {\bibfnamefont {J.}~\bibnamefont {Danon}},
  \bibinfo {author} {\bibfnamefont {M.}~\bibnamefont {Leijnse}}, \bibinfo
  {author} {\bibfnamefont {K.}~\bibnamefont {Flensberg}}, \bibinfo {author}
  {\bibfnamefont {J.}~\bibnamefont {Nyg{\aa}rd}}, \bibinfo {author}
  {\bibfnamefont {P.}~\bibnamefont {Krogstrup}}, \ and\ \bibinfo {author}
  {\bibfnamefont {C.~M.}\ \bibnamefont {Marcus}},\ }\bibfield  {title}
  {\enquote {\bibinfo {title} {{Majorana} bound state in a coupled quantum-dot
  hybrid-nanowire system}}, }\href {https://doi.org/10.1126/science.aaf3961}
  {\bibfield  {journal} {\bibinfo  {journal} {Science}\ }\textbf {\bibinfo
  {volume} {354}},\ \bibinfo {pages} {1557} (\bibinfo {year}
  {2016})}\BibitemShut {NoStop}%
\bibitem [{\citenamefont {Chen}\ \emph {et~al.}(2017)\citenamefont {Chen},
  \citenamefont {Yu}, \citenamefont {Stenger}, \citenamefont {Hocevar},
  \citenamefont {Car}, \citenamefont {Plissard}, \citenamefont {Bakkers},
  \citenamefont {Stanescu},\ and\ \citenamefont
  {Frolov}}]{chen2017experimental}%
  \BibitemOpen
  \bibfield  {author} {\bibinfo {author} {\bibfnamefont {J.}~\bibnamefont
  {Chen}}, \bibinfo {author} {\bibfnamefont {P.}~\bibnamefont {Yu}}, \bibinfo
  {author} {\bibfnamefont {J.}~\bibnamefont {Stenger}}, \bibinfo {author}
  {\bibfnamefont {M.}~\bibnamefont {Hocevar}}, \bibinfo {author} {\bibfnamefont
  {D.}~\bibnamefont {Car}}, \bibinfo {author} {\bibfnamefont {S.~R.}\
  \bibnamefont {Plissard}}, \bibinfo {author} {\bibfnamefont {E.~P.}\
  \bibnamefont {Bakkers}}, \bibinfo {author} {\bibfnamefont {T.~D.}\
  \bibnamefont {Stanescu}}, \ and\ \bibinfo {author} {\bibfnamefont {S.~M.}\
  \bibnamefont {Frolov}},\ }\bibfield  {title} {\enquote {\bibinfo {title}
  {Experimental phase diagram of zero-bias conductance peaks in
  superconductor/semiconductor nanowire devices}}, }\href
  {https://doi.org/10.1126/sciadv.1701476} {\bibfield  {journal} {\bibinfo
  {journal} {Sci. Adv.}\ }\textbf {\bibinfo {volume} {3}},\ \bibinfo {pages}
  {e1701476} (\bibinfo {year} {2017})}\BibitemShut {NoStop}%
\bibitem [{\citenamefont {Suominen}\ \emph {et~al.}(2017)\citenamefont
  {Suominen}, \citenamefont {Kjaergaard}, \citenamefont {Hamilton},
  \citenamefont {Shabani}, \citenamefont {Palmstr{\o}m}, \citenamefont
  {Marcus},\ and\ \citenamefont {Nichele}}]{suominen2017zero}%
  \BibitemOpen
  \bibfield  {author} {\bibinfo {author} {\bibfnamefont {H.~J.}\ \bibnamefont
  {Suominen}}, \bibinfo {author} {\bibfnamefont {M.}~\bibnamefont
  {Kjaergaard}}, \bibinfo {author} {\bibfnamefont {A.~R.}\ \bibnamefont
  {Hamilton}}, \bibinfo {author} {\bibfnamefont {J.}~\bibnamefont {Shabani}},
  \bibinfo {author} {\bibfnamefont {C.~J.}\ \bibnamefont {Palmstr{\o}m}},
  \bibinfo {author} {\bibfnamefont {C.~M.}\ \bibnamefont {Marcus}}, \ and\
  \bibinfo {author} {\bibfnamefont {F.}~\bibnamefont {Nichele}},\ }\bibfield
  {title} {\enquote {\bibinfo {title} {Zero-energy modes from coalescing
  {Andreev} states in a two-dimensional semiconductor-superconductor hybrid
  platform}}, }\href {https://doi.org/10.1103/PhysRevLett.119.176805}
  {\bibfield  {journal} {\bibinfo  {journal} {Phys. Rev. Lett.}\ }\textbf
  {\bibinfo {volume} {119}},\ \bibinfo {pages} {176805} (\bibinfo {year}
  {2017})}\BibitemShut {NoStop}%
\bibitem [{\citenamefont {Nichele}\ \emph {et~al.}(2017)\citenamefont
  {Nichele}, \citenamefont {Drachmann}, \citenamefont {Whiticar}, \citenamefont
  {O¡¯Farrell}, \citenamefont {Suominen}, \citenamefont {Fornieri},
  \citenamefont {Wang}, \citenamefont {Gardner}, \citenamefont {Thomas},
  \citenamefont {Hatke} \emph {et~al.}}]{nichele2017scaling}%
  \BibitemOpen
  \bibfield  {author} {\bibinfo {author} {\bibfnamefont {F.}~\bibnamefont
  {Nichele}},  \emph {et~al.},\ }\bibfield  {title} {\enquote {\bibinfo {title}
  {Scaling of {Majorana} zero-bias conductance peaks}}, }\href
  {https://doi.org/10.1103/PhysRevLett.119.136803} {\bibfield  {journal}
  {\bibinfo  {journal} {Phys. Rev. Lett.}\ }\textbf {\bibinfo {volume} {119}},\
  \bibinfo {pages} {136803} (\bibinfo {year} {2017})}\BibitemShut {NoStop}%
\bibitem [{\citenamefont {Zhang}\ \emph {et~al.}(2018)\citenamefont {Zhang},
  \citenamefont {Liu}, \citenamefont {Gazibegovic}, \citenamefont {Xu},
  \citenamefont {Logan}, \citenamefont {Wang}, \citenamefont {Van~Loo},
  \citenamefont {Bommer}, \citenamefont {De~Moor}, \citenamefont {Car} \emph
  {et~al.}}]{zhang2018quantized}%
  \BibitemOpen
  \bibfield  {author} {\bibinfo {author} {\bibfnamefont {H.}~\bibnamefont
  {Zhang}},  \emph {et~al.},\ }\bibfield  {title} {\enquote {\bibinfo {title}
  {Quantized {Majorana} conductance}}, }\href
  {https://doi.org/10.1038/nature26142} {\bibfield  {journal} {\bibinfo
  {journal} {Nature (London)}\ }\textbf {\bibinfo {volume} {556}},\ \bibinfo
  {pages} {74} (\bibinfo {year} {2018})}\BibitemShut {NoStop}%
\bibitem [{\citenamefont {G{\"u}l}\ \emph {et~al.}(2018)\citenamefont
  {G{\"u}l}, \citenamefont {Zhang}, \citenamefont {Bommer}, \citenamefont
  {de~Moor}, \citenamefont {Car}, \citenamefont {Plissard}, \citenamefont
  {Bakkers}, \citenamefont {Geresdi}, \citenamefont {Watanabe}, \citenamefont
  {Taniguchi} \emph {et~al.}}]{gul2018ballistic}%
  \BibitemOpen
  \bibfield  {author} {\bibinfo {author} {\bibfnamefont {{\"O}.}~\bibnamefont
  {G{\"u}l}},  \emph {et~al.},\ }\bibfield  {title} {\enquote {\bibinfo {title}
  {Ballistic {Majorana} nanowire devices}}, }\href
  {https://doi.org/10.1038/s41565-017-0032-8} {\bibfield  {journal} {\bibinfo
  {journal} {Nat. Nanotechnol.}\ }\textbf {\bibinfo {volume} {13}},\ \bibinfo
  {pages} {192} (\bibinfo {year} {2018})}\BibitemShut {NoStop}%
\bibitem [{\citenamefont {Sestoft}\ \emph {et~al.}(2018)\citenamefont
  {Sestoft}, \citenamefont {Kanne}, \citenamefont {Gejl}, \citenamefont {von
  Soosten}, \citenamefont {Yodh}, \citenamefont {Sherman}, \citenamefont
  {Tarasinski}, \citenamefont {Wimmer}, \citenamefont {Johnson}, \citenamefont
  {Deng} \emph {et~al.}}]{sestoft2018engineering}%
  \BibitemOpen
  \bibfield  {author} {\bibinfo {author} {\bibfnamefont {J.~E.}\ \bibnamefont
  {Sestoft}},  \emph {et~al.},\ }\bibfield  {title} {\enquote {\bibinfo {title}
  {Engineering hybrid epitaxial {InAsSb/Al} nanowires for stronger topological
  protection}}, }\href {https://doi.org/10.1103/PhysRevMaterials.2.044202}
  {\bibfield  {journal} {\bibinfo  {journal} {Phys. Rev. Mater.}\ }\textbf
  {\bibinfo {volume} {2}},\ \bibinfo {pages} {044202} (\bibinfo {year}
  {2018})}\BibitemShut {NoStop}%
\bibitem [{\citenamefont {Vaitiek{\.e}nas}\ \emph {et~al.}(2018)\citenamefont
  {Vaitiek{\.e}nas}, \citenamefont {Deng}, \citenamefont {Nyg\aa{}rd},
  \citenamefont {Krogstrup},\ and\ \citenamefont
  {Marcus}}]{vaitiekenas2018effective}%
  \BibitemOpen
  \bibfield  {author} {\bibinfo {author} {\bibfnamefont {S.}~\bibnamefont
  {Vaitiek{\.e}nas}}, \bibinfo {author} {\bibfnamefont {M.-T.}\ \bibnamefont
  {Deng}}, \bibinfo {author} {\bibfnamefont {J.}~\bibnamefont {Nyg\aa{}rd}},
  \bibinfo {author} {\bibfnamefont {P.}~\bibnamefont {Krogstrup}}, \ and\
  \bibinfo {author} {\bibfnamefont {C.~M.}\ \bibnamefont {Marcus}},\ }\bibfield
   {title} {\enquote {\bibinfo {title} {Effective $g$ factor of subgap states
  in hybrid nanowires}}, }\href {\doibase 10.1103/PhysRevLett.121.037703}
  {\bibfield  {journal} {\bibinfo  {journal} {Phys. Rev. Lett.}\ }\textbf
  {\bibinfo {volume} {121}},\ \bibinfo {pages} {037703} (\bibinfo {year}
  {2018})}\BibitemShut {NoStop}%
\bibitem [{\citenamefont {Deng}\ \emph {et~al.}(2018)\citenamefont {Deng},
  \citenamefont {Vaitiek{\.e}nas}, \citenamefont {Prada}, \citenamefont
  {San-Jose}, \citenamefont {Nyg\aa{}rd}, \citenamefont {Krogstrup},
  \citenamefont {Aguado},\ and\ \citenamefont {Marcus}}]{deng2018nonlocality}%
  \BibitemOpen
  \bibfield  {author} {\bibinfo {author} {\bibfnamefont {M.-T.}\ \bibnamefont
  {Deng}}, \bibinfo {author} {\bibfnamefont {S.}~\bibnamefont
  {Vaitiek{\.e}nas}}, \bibinfo {author} {\bibfnamefont {E.}~\bibnamefont
  {Prada}}, \bibinfo {author} {\bibfnamefont {P.}~\bibnamefont {San-Jose}},
  \bibinfo {author} {\bibfnamefont {J.}~\bibnamefont {Nyg\aa{}rd}}, \bibinfo
  {author} {\bibfnamefont {P.}~\bibnamefont {Krogstrup}}, \bibinfo {author}
  {\bibfnamefont {R.}~\bibnamefont {Aguado}}, \ and\ \bibinfo {author}
  {\bibfnamefont {C.~M.}\ \bibnamefont {Marcus}},\ }\bibfield  {title}
  {\enquote {\bibinfo {title} {Nonlocality of {Majorana} modes in hybrid
  nanowires}}, }\href {\doibase 10.1103/PhysRevB.98.085125} {\bibfield
  {journal} {\bibinfo  {journal} {Phys. Rev. B}\ }\textbf {\bibinfo {volume}
  {98}},\ \bibinfo {pages} {085125} (\bibinfo {year} {2018})}\BibitemShut
  {NoStop}%
\bibitem [{\citenamefont {de~Moor}\ \emph {et~al.}(2018)\citenamefont
  {de~Moor}, \citenamefont {Bommer}, \citenamefont {Xu}, \citenamefont
  {Winkler}, \citenamefont {Antipov}, \citenamefont {Bargerbos}, \citenamefont
  {Wang}, \citenamefont {van Loo}, \citenamefont {Veld}, \citenamefont
  {Gazibegovic} \emph {et~al.}}]{de2018electric}%
  \BibitemOpen
  \bibfield  {author} {\bibinfo {author} {\bibfnamefont {M.~W.}\ \bibnamefont
  {de~Moor}},  \emph {et~al.},\ }\bibfield  {title} {\enquote {\bibinfo {title}
  {Electric field tunable superconductor-semiconductor coupling in {Majorana}
  nanowires}}, }\href {https://doi.org/10.1088/1367-2630/aae61d} {\bibfield
  {journal} {\bibinfo  {journal} {New J. Phys.}\ }\textbf {\bibinfo {volume}
  {20}},\ \bibinfo {pages} {103049} (\bibinfo {year} {2018})}\BibitemShut
  {NoStop}%
\bibitem [{\citenamefont {Bommer}\ \emph {et~al.}(2018)\citenamefont {Bommer},
  \citenamefont {Zhang}, \citenamefont {G{\"u}l}, \citenamefont {Nijholt},
  \citenamefont {Wimmer}, \citenamefont {Rybakov}, \citenamefont {Garaud},
  \citenamefont {Rodic}, \citenamefont {Babaev}, \citenamefont {Troyer} \emph
  {et~al.}}]{bommer2018spin}%
  \BibitemOpen
  \bibfield  {author} {\bibinfo {author} {\bibfnamefont {J.~D.}\ \bibnamefont
  {Bommer}},  \emph {et~al.},\ }\bibfield  {title} {\enquote {\bibinfo {title}
  {Spin-orbit protection of induced superconductivity in {Majorana}
  nanowires}}, }\href {https://arxiv.org/abs/1807.01940} {\bibfield  {journal}
  {\bibinfo  {journal} {arXiv:1807.01940}\ } (\bibinfo {year}
  {2018})}\BibitemShut {NoStop}%
\bibitem [{\citenamefont {Lutchyn}\ \emph {et~al.}(2018)\citenamefont
  {Lutchyn}, \citenamefont {Bakkers}, \citenamefont {Kouwenhoven},
  \citenamefont {Krogstrup}, \citenamefont {Marcus},\ and\ \citenamefont
  {Oreg}}]{lutchyn2018majorana}%
  \BibitemOpen
  \bibfield  {author} {\bibinfo {author} {\bibfnamefont {R.}~\bibnamefont
  {Lutchyn}}, \bibinfo {author} {\bibfnamefont {E.}~\bibnamefont {Bakkers}},
  \bibinfo {author} {\bibfnamefont {L.}~\bibnamefont {Kouwenhoven}}, \bibinfo
  {author} {\bibfnamefont {P.}~\bibnamefont {Krogstrup}}, \bibinfo {author}
  {\bibfnamefont {C.}~\bibnamefont {Marcus}}, \ and\ \bibinfo {author}
  {\bibfnamefont {Y.}~\bibnamefont {Oreg}},\ }\bibfield  {title} {\enquote
  {\bibinfo {title} {{Majorana} zero modes in superconductor--semiconductor
  heterostructures}}, }\href {https://doi.org/10.1038/s41578-018-0003-1}
  {\bibfield  {journal} {\bibinfo  {journal} {Nat. Rev. Mater.}\ }\textbf
  {\bibinfo {volume} {3}},\ \bibinfo {pages} {52} (\bibinfo {year}
  {2018})}\BibitemShut {NoStop}%
\bibitem [{\citenamefont {Prada}\ \emph {et~al.}(2012)\citenamefont {Prada},
  \citenamefont {San-Jose},\ and\ \citenamefont {Aguado}}]{prada2012transport}%
  \BibitemOpen
  \bibfield  {author} {\bibinfo {author} {\bibfnamefont {E.}~\bibnamefont
  {Prada}}, \bibinfo {author} {\bibfnamefont {P.}~\bibnamefont {San-Jose}}, \
  and\ \bibinfo {author} {\bibfnamefont {R.}~\bibnamefont {Aguado}},\
  }\bibfield  {title} {\enquote {\bibinfo {title} {Transport spectroscopy of
  {$NS$} nanowire junctions with {Majorana} fermions}}, }\href
  {https://doi.org/10.1103/PhysRevB.86.180503} {\bibfield  {journal} {\bibinfo
  {journal} {Phys. Rev. B}\ }\textbf {\bibinfo {volume} {86}},\ \bibinfo
  {pages} {180503(R)} (\bibinfo {year} {2012})}\BibitemShut {NoStop}%
\bibitem [{\citenamefont {{Das Sarma}}\ \emph {et~al.}(2012)\citenamefont {{Das
  Sarma}}, \citenamefont {Sau},\ and\ \citenamefont
  {Stanescu}}]{sarma2012splitting}%
  \BibitemOpen
  \bibfield  {author} {\bibinfo {author} {\bibfnamefont {S.}~\bibnamefont {{Das
  Sarma}}}, \bibinfo {author} {\bibfnamefont {J.~D.}\ \bibnamefont {Sau}}, \
  and\ \bibinfo {author} {\bibfnamefont {T.~D.}\ \bibnamefont {Stanescu}},\
  }\bibfield  {title} {\enquote {\bibinfo {title} {Splitting of the zero-bias
  conductance peak as smoking gun evidence for the existence of the {Majorana}
  mode in a superconductor-semiconductor nanowire}}, }\href
  {https://doi.org/10.1103/PhysRevB.86.220506} {\bibfield  {journal} {\bibinfo
  {journal} {Phys. Rev. B}\ }\textbf {\bibinfo {volume} {86}},\ \bibinfo
  {pages} {220506(R)} (\bibinfo {year} {2012})}\BibitemShut {NoStop}%
\bibitem [{\citenamefont {Rainis}\ \emph {et~al.}(2013)\citenamefont {Rainis},
  \citenamefont {Trifunovic}, \citenamefont {Klinovaja},\ and\ \citenamefont
  {Loss}}]{rainis2013towards}%
  \BibitemOpen
  \bibfield  {author} {\bibinfo {author} {\bibfnamefont {D.}~\bibnamefont
  {Rainis}}, \bibinfo {author} {\bibfnamefont {L.}~\bibnamefont {Trifunovic}},
  \bibinfo {author} {\bibfnamefont {J.}~\bibnamefont {Klinovaja}}, \ and\
  \bibinfo {author} {\bibfnamefont {D.}~\bibnamefont {Loss}},\ }\bibfield
  {title} {\enquote {\bibinfo {title} {Towards a realistic transport modeling
  in a superconducting nanowire with {Majorana} fermions}}, }\href
  {https://doi.org/10.1103/PhysRevB.87.024515} {\bibfield  {journal} {\bibinfo
  {journal} {Phys. Rev. B}\ }\textbf {\bibinfo {volume} {87}},\ \bibinfo
  {pages} {024515} (\bibinfo {year} {2013})}\BibitemShut {NoStop}%
\bibitem [{\citenamefont {Albrecht}\ \emph {et~al.}(2016)\citenamefont
  {Albrecht}, \citenamefont {Higginbotham}, \citenamefont {Madsen},
  \citenamefont {Kuemmeth}, \citenamefont {Jespersen}, \citenamefont
  {Nyg{\aa}rd}, \citenamefont {Krogstrup},\ and\ \citenamefont
  {Marcus}}]{albrecht2016exponential}%
  \BibitemOpen
  \bibfield  {author} {\bibinfo {author} {\bibfnamefont {S.~M.}\ \bibnamefont
  {Albrecht}}, \bibinfo {author} {\bibfnamefont {A.}~\bibnamefont
  {Higginbotham}}, \bibinfo {author} {\bibfnamefont {M.}~\bibnamefont
  {Madsen}}, \bibinfo {author} {\bibfnamefont {F.}~\bibnamefont {Kuemmeth}},
  \bibinfo {author} {\bibfnamefont {T.~S.}\ \bibnamefont {Jespersen}}, \bibinfo
  {author} {\bibfnamefont {J.}~\bibnamefont {Nyg{\aa}rd}}, \bibinfo {author}
  {\bibfnamefont {P.}~\bibnamefont {Krogstrup}}, \ and\ \bibinfo {author}
  {\bibfnamefont {C.}~\bibnamefont {Marcus}},\ }\bibfield  {title} {\enquote
  {\bibinfo {title} {Exponential protection of zero modes in {Majorana}
  islands}}, }\href {https://doi.org/10.1038/nature17162} {\bibfield  {journal}
  {\bibinfo  {journal} {Nature (London)}\ }\textbf {\bibinfo {volume} {531}},\
  \bibinfo {pages} {206} (\bibinfo {year} {2016})}\BibitemShut {NoStop}%
\bibitem [{\citenamefont {Sherman}\ \emph {et~al.}(2017)\citenamefont
  {Sherman}, \citenamefont {Yodh}, \citenamefont {Albrecht}, \citenamefont
  {Nyg{\aa}rd}, \citenamefont {Krogstrup},\ and\ \citenamefont
  {Marcus}}]{sherman2017normal}%
  \BibitemOpen
  \bibfield  {author} {\bibinfo {author} {\bibfnamefont {D.}~\bibnamefont
  {Sherman}}, \bibinfo {author} {\bibfnamefont {J.}~\bibnamefont {Yodh}},
  \bibinfo {author} {\bibfnamefont {S.}~\bibnamefont {Albrecht}}, \bibinfo
  {author} {\bibfnamefont {J.}~\bibnamefont {Nyg{\aa}rd}}, \bibinfo {author}
  {\bibfnamefont {P.}~\bibnamefont {Krogstrup}}, \ and\ \bibinfo {author}
  {\bibfnamefont {C.}~\bibnamefont {Marcus}},\ }\bibfield  {title} {\enquote
  {\bibinfo {title} {Normal, superconducting and topological regimes of hybrid
  double quantum dots}}, }\href {https://doi.org/10.1038/nnano.2016.227}
  {\bibfield  {journal} {\bibinfo  {journal} {Nat. Nanotechnol.}\ }\textbf
  {\bibinfo {volume} {12}},\ \bibinfo {pages} {212} (\bibinfo {year}
  {2017})}\BibitemShut {NoStop}%
\bibitem [{\citenamefont {Albrecht}\ \emph {et~al.}(2017)\citenamefont
  {Albrecht}, \citenamefont {Hansen}, \citenamefont {Higginbotham},
  \citenamefont {Kuemmeth}, \citenamefont {Jespersen}, \citenamefont
  {Nyg\aa{}rd}, \citenamefont {Krogstrup}, \citenamefont {Danon}, \citenamefont
  {Flensberg},\ and\ \citenamefont {Marcus}}]{albrecht2017transport}%
  \BibitemOpen
  \bibfield  {author} {\bibinfo {author} {\bibfnamefont {S.~M.}\ \bibnamefont
  {Albrecht}}, \bibinfo {author} {\bibfnamefont {E.~B.}\ \bibnamefont
  {Hansen}}, \bibinfo {author} {\bibfnamefont {A.~P.}\ \bibnamefont
  {Higginbotham}}, \bibinfo {author} {\bibfnamefont {F.}~\bibnamefont
  {Kuemmeth}}, \bibinfo {author} {\bibfnamefont {T.~S.}\ \bibnamefont
  {Jespersen}}, \bibinfo {author} {\bibfnamefont {J.}~\bibnamefont
  {Nyg\aa{}rd}}, \bibinfo {author} {\bibfnamefont {P.}~\bibnamefont
  {Krogstrup}}, \bibinfo {author} {\bibfnamefont {J.}~\bibnamefont {Danon}},
  \bibinfo {author} {\bibfnamefont {K.}~\bibnamefont {Flensberg}}, \ and\
  \bibinfo {author} {\bibfnamefont {C.~M.}\ \bibnamefont {Marcus}},\ }\bibfield
   {title} {\enquote {\bibinfo {title} {Transport signatures of quasiparticle
  poisoning in a {Majorana} island}}, }\href {\doibase
  10.1103/PhysRevLett.118.137701} {\bibfield  {journal} {\bibinfo  {journal}
  {Phys. Rev. Lett.}\ }\textbf {\bibinfo {volume} {118}},\ \bibinfo {pages}
  {137701} (\bibinfo {year} {2017})}\BibitemShut {NoStop}%
\bibitem [{\citenamefont {Vaitiek\ifmmode~\dot{e}\else \.{e}\fi{}nas}\ \emph
  {et~al.}(2018)\citenamefont {Vaitiek\ifmmode~\dot{e}\else \.{e}\fi{}nas},
  \citenamefont {Whiticar}, \citenamefont {Deng}, \citenamefont {Krizek},
  \citenamefont {Sestoft}, \citenamefont {Palmstr\o{}m}, \citenamefont
  {Marti-Sanchez}, \citenamefont {Arbiol}, \citenamefont {Krogstrup},
  \citenamefont {Casparis},\ and\ \citenamefont
  {Marcus}}]{vaitiekenas2018selective}%
  \BibitemOpen
  \bibfield  {author} {\bibinfo {author} {\bibfnamefont {S.}~\bibnamefont
  {Vaitiek\ifmmode~\dot{e}\else \.{e}\fi{}nas}},  \emph {et~al.},\ }\bibfield
  {title} {\enquote {\bibinfo {title} {Selective-area-grown
  semiconductor-superconductor hybrids: A basis for topological networks}},
  }\href {\doibase 10.1103/PhysRevLett.121.147701} {\bibfield  {journal}
  {\bibinfo  {journal} {Phys. Rev. Lett.}\ }\textbf {\bibinfo {volume} {121}},\
  \bibinfo {pages} {147701} (\bibinfo {year} {2018})}\BibitemShut {NoStop}%
\bibitem [{\citenamefont {O'Farrell}\ \emph {et~al.}(2018)\citenamefont
  {O'Farrell}, \citenamefont {Drachmann}, \citenamefont {Hell}, \citenamefont
  {Fornieri}, \citenamefont {Whiticar}, \citenamefont {Hansen}, \citenamefont
  {Gronin}, \citenamefont {Gardner}, \citenamefont {Thomas}, \citenamefont
  {Manfra}, \citenamefont {Flensberg}, \citenamefont {Marcus},\ and\
  \citenamefont {Nichele}}]{o2018hybridization}%
  \BibitemOpen
  \bibfield  {author} {\bibinfo {author} {\bibfnamefont {E.~C.~T.}\
  \bibnamefont {O'Farrell}},  \emph {et~al.},\ }\bibfield  {title} {\enquote
  {\bibinfo {title} {Hybridization of subgap states in one-dimensional
  superconductor-semiconductor {Coulomb} islands}}, }\href {\doibase
  10.1103/PhysRevLett.121.256803} {\bibfield  {journal} {\bibinfo  {journal}
  {Phys. Rev. Lett.}\ }\textbf {\bibinfo {volume} {121}},\ \bibinfo {pages}
  {256803} (\bibinfo {year} {2018})}\BibitemShut {NoStop}%
\bibitem [{\citenamefont {Shen}\ \emph {et~al.}(2018)\citenamefont {Shen},
  \citenamefont {Heedt}, \citenamefont {Borsoi}, \citenamefont {Van~Heck},
  \citenamefont {Gazibegovic}, \citenamefont {Veld}, \citenamefont {Car},
  \citenamefont {Logan}, \citenamefont {Pendharkar}, \citenamefont {Wang} \emph
  {et~al.}}]{shen2018parity}%
  \BibitemOpen
  \bibfield  {author} {\bibinfo {author} {\bibfnamefont {J.}~\bibnamefont
  {Shen}},  \emph {et~al.},\ }\bibfield  {title} {\enquote {\bibinfo {title}
  {Parity transitions in the superconducting ground state of hybrid {InSb-Al}
  {Coulomb} islands}}, }\href {https://doi.org/10.1038/s41467-018-07279-7}
  {\bibfield  {journal} {\bibinfo  {journal} {Nat. Commun.}\ }\textbf {\bibinfo
  {volume} {9}},\ \bibinfo {pages} {4801} (\bibinfo {year} {2018})}\BibitemShut
  {NoStop}%
\bibitem [{\citenamefont {Chiu}\ \emph {et~al.}(2017)\citenamefont {Chiu},
  \citenamefont {Sau},\ and\ \citenamefont {{Das
  Sarma}}}]{chiu2017conductance}%
  \BibitemOpen
  \bibfield  {author} {\bibinfo {author} {\bibfnamefont {C.-K.}\ \bibnamefont
  {Chiu}}, \bibinfo {author} {\bibfnamefont {J.~D.}\ \bibnamefont {Sau}}, \
  and\ \bibinfo {author} {\bibfnamefont {S.}~\bibnamefont {{Das Sarma}}},\
  }\bibfield  {title} {\enquote {\bibinfo {title} {Conductance of a
  superconducting {Coulomb}-blockaded {Majorana} nanowire}}, }\href
  {https://doi.org/10.1103/PhysRevB.96.054504} {\bibfield  {journal} {\bibinfo
  {journal} {Phys. Rev. B}\ }\textbf {\bibinfo {volume} {96}},\ \bibinfo
  {pages} {054504} (\bibinfo {year} {2017})}\BibitemShut {NoStop}%
\bibitem [{\citenamefont {Dmytruk}\ and\ \citenamefont
  {Klinovaja}(2018)}]{dmytruk2018suppression}%
  \BibitemOpen
  \bibfield  {author} {\bibinfo {author} {\bibfnamefont {O.}~\bibnamefont
  {Dmytruk}}\ and\ \bibinfo {author} {\bibfnamefont {J.}~\bibnamefont
  {Klinovaja}},\ }\bibfield  {title} {\enquote {\bibinfo {title} {Suppression
  of the overlap between {Majorana} fermions by orbital magnetic effects in
  semiconducting-superconducting nanowires}}, }\href
  {https://doi.org/10.1103/PhysRevB.97.155409} {\bibfield  {journal} {\bibinfo
  {journal} {Phys. Rev. B}\ }\textbf {\bibinfo {volume} {97}},\ \bibinfo
  {pages} {155409} (\bibinfo {year} {2018})}\BibitemShut {NoStop}%
\bibitem [{\citenamefont {Fleckenstein}\ \emph {et~al.}(2018)\citenamefont
  {Fleckenstein}, \citenamefont {Dom\'{\i}nguez}, \citenamefont
  {Traverso~Ziani},\ and\ \citenamefont
  {Trauzettel}}]{fleckenstein2018decaying}%
  \BibitemOpen
  \bibfield  {author} {\bibinfo {author} {\bibfnamefont {C.}~\bibnamefont
  {Fleckenstein}}, \bibinfo {author} {\bibfnamefont {F.}~\bibnamefont
  {Dom\'{\i}nguez}}, \bibinfo {author} {\bibfnamefont {N.}~\bibnamefont
  {Traverso~Ziani}}, \ and\ \bibinfo {author} {\bibfnamefont {B.}~\bibnamefont
  {Trauzettel}},\ }\bibfield  {title} {\enquote {\bibinfo {title} {Decaying
  spectral oscillations in a {Majorana} wire with finite coherence length}},
  }\href {\doibase 10.1103/PhysRevB.97.155425} {\bibfield  {journal} {\bibinfo
  {journal} {Phys. Rev. B}\ }\textbf {\bibinfo {volume} {97}},\ \bibinfo
  {pages} {155425} (\bibinfo {year} {2018})}\BibitemShut {NoStop}%
\bibitem [{\citenamefont {Plugge}\ \emph {et~al.}(2017)\citenamefont {Plugge},
  \citenamefont {Rasmussen}, \citenamefont {Egger},\ and\ \citenamefont
  {Flensberg}}]{plugge2017majorana}%
  \BibitemOpen
  \bibfield  {author} {\bibinfo {author} {\bibfnamefont {S.}~\bibnamefont
  {Plugge}}, \bibinfo {author} {\bibfnamefont {A.}~\bibnamefont {Rasmussen}},
  \bibinfo {author} {\bibfnamefont {R.}~\bibnamefont {Egger}}, \ and\ \bibinfo
  {author} {\bibfnamefont {K.}~\bibnamefont {Flensberg}},\ }\bibfield  {title}
  {\enquote {\bibinfo {title} {{Majorana} box qubits}}, }\href
  {https://doi.org/10.1088/1367-2630/aa54e1} {\bibfield  {journal} {\bibinfo
  {journal} {New J. Phys.}\ }\textbf {\bibinfo {volume} {19}},\ \bibinfo
  {pages} {012001} (\bibinfo {year} {2017})}\BibitemShut {NoStop}%
\bibitem [{\citenamefont {Karzig}\ \emph {et~al.}(2017)\citenamefont {Karzig},
  \citenamefont {Knapp}, \citenamefont {Lutchyn}, \citenamefont {Bonderson},
  \citenamefont {Hastings}, \citenamefont {Nayak}, \citenamefont {Alicea},
  \citenamefont {Flensberg}, \citenamefont {Plugge}, \citenamefont {Oreg} \emph
  {et~al.}}]{karzig2017scalable}%
  \BibitemOpen
  \bibfield  {author} {\bibinfo {author} {\bibfnamefont {T.}~\bibnamefont
  {Karzig}},  \emph {et~al.},\ }\bibfield  {title} {\enquote {\bibinfo {title}
  {Scalable designs for quasiparticle-poisoning-protected topological quantum
  computation with {Majorana} zero modes}}, }\href
  {https://doi.org/10.1103/PhysRevB.95.235305} {\bibfield  {journal} {\bibinfo
  {journal} {Phys. Rev. B}\ }\textbf {\bibinfo {volume} {95}},\ \bibinfo
  {pages} {235305} (\bibinfo {year} {2017})}\BibitemShut {NoStop}%
\bibitem [{\citenamefont {Winkler}(2003)}]{winkler2003spin}%
  \BibitemOpen
  \bibfield  {author} {\bibinfo {author} {\bibfnamefont {R.}~\bibnamefont
  {Winkler}},\ }\href@noop {} {\emph {\bibinfo {title} {Spin-orbit coupling
  effects in two-dimensional electron and hole systems}}}\ (\bibinfo
  {publisher} {Springer Science \& Business Media, Berlin},\ \bibinfo {year}
  {2003})\BibitemShut {NoStop}%
\bibitem [{\citenamefont {S{\'a}nchez}\ and\ \citenamefont
  {Serra}(2006)}]{sanchez2006fano}%
  \BibitemOpen
  \bibfield  {author} {\bibinfo {author} {\bibfnamefont {D.}~\bibnamefont
  {S{\'a}nchez}}\ and\ \bibinfo {author} {\bibfnamefont {L.}~\bibnamefont
  {Serra}},\ }\bibfield  {title} {\enquote {\bibinfo {title} {Fano-{Rashba}
  effect in a quantum wire}}, }\href
  {https://doi.org/10.1103/PhysRevB.74.153313} {\bibfield  {journal} {\bibinfo
  {journal} {Phys. Rev. B}\ }\textbf {\bibinfo {volume} {74}},\ \bibinfo
  {pages} {153313} (\bibinfo {year} {2006})}\BibitemShut {NoStop}%
\bibitem [{\citenamefont {S{\'a}nchez}\ \emph {et~al.}(2008)\citenamefont
  {S{\'a}nchez}, \citenamefont {Serra},\ and\ \citenamefont
  {Choi}}]{sanchez2008strongly}%
  \BibitemOpen
  \bibfield  {author} {\bibinfo {author} {\bibfnamefont {D.}~\bibnamefont
  {S{\'a}nchez}}, \bibinfo {author} {\bibfnamefont {L.}~\bibnamefont {Serra}},
  \ and\ \bibinfo {author} {\bibfnamefont {M.-S.}\ \bibnamefont {Choi}},\
  }\bibfield  {title} {\enquote {\bibinfo {title} {Strongly modulated
  transmission of a spin-split quantum wire with local {Rashba} interaction}},
  }\href {https://doi.org/10.1103/PhysRevB.77.035315} {\bibfield  {journal}
  {\bibinfo  {journal} {Phys. Rev. B}\ }\textbf {\bibinfo {volume} {77}},\
  \bibinfo {pages} {035315} (\bibinfo {year} {2008})}\BibitemShut {NoStop}%
\bibitem [{\citenamefont {Glazov}\ and\ \citenamefont
  {Sherman}(2011)}]{glazov2011theory}%
  \BibitemOpen
  \bibfield  {author} {\bibinfo {author} {\bibfnamefont {M.~M.}\ \bibnamefont
  {Glazov}}\ and\ \bibinfo {author} {\bibfnamefont {E.~Y.}\ \bibnamefont
  {Sherman}},\ }\bibfield  {title} {\enquote {\bibinfo {title} {Theory of spin
  noise in nanowires}}, }\href {\doibase 10.1103/PhysRevLett.107.156602}
  {\bibfield  {journal} {\bibinfo  {journal} {Phys. Rev. Lett.}\ }\textbf
  {\bibinfo {volume} {107}},\ \bibinfo {pages} {156602} (\bibinfo {year}
  {2011})}\BibitemShut {NoStop}%
\bibitem [{\citenamefont {Sadreev}\ and\ \citenamefont
  {Sherman}(2013)}]{sadreev2013effect}%
  \BibitemOpen
  \bibfield  {author} {\bibinfo {author} {\bibfnamefont {A.~F.}\ \bibnamefont
  {Sadreev}}\ and\ \bibinfo {author} {\bibfnamefont {E.~Y.}\ \bibnamefont
  {Sherman}},\ }\bibfield  {title} {\enquote {\bibinfo {title} {Effect of
  gate-driven spin resonance on the conductance through a one-dimensional
  quantum wire}}, }\href {https://doi.org/10.1103/PhysRevB.88.115302}
  {\bibfield  {journal} {\bibinfo  {journal} {Phys. Rev. B}\ }\textbf {\bibinfo
  {volume} {88}},\ \bibinfo {pages} {115302} (\bibinfo {year}
  {2013})}\BibitemShut {NoStop}%
\bibitem [{\citenamefont {Modugno}\ \emph {et~al.}(2017)\citenamefont
  {Modugno}, \citenamefont {Sherman},\ and\ \citenamefont
  {Konotop}}]{modugno2017macroscopic}%
  \BibitemOpen
  \bibfield  {author} {\bibinfo {author} {\bibfnamefont {M.}~\bibnamefont
  {Modugno}}, \bibinfo {author} {\bibfnamefont {E.~Y.}\ \bibnamefont
  {Sherman}}, \ and\ \bibinfo {author} {\bibfnamefont {V.~V.}\ \bibnamefont
  {Konotop}},\ }\bibfield  {title} {\enquote {\bibinfo {title} {Macroscopic
  random {Paschen-Back} effect in ultracold atomic gases}}, }\href {\doibase
  10.1103/PhysRevA.95.063620} {\bibfield  {journal} {\bibinfo  {journal} {Phys.
  Rev. A}\ }\textbf {\bibinfo {volume} {95}},\ \bibinfo {pages} {063620}
  (\bibinfo {year} {2017})}\BibitemShut {NoStop}%
\bibitem [{\citenamefont {Klinovaja}\ and\ \citenamefont
  {Loss}(2015)}]{klinovaja2015fermionic}%
  \BibitemOpen
  \bibfield  {author} {\bibinfo {author} {\bibfnamefont {J.}~\bibnamefont
  {Klinovaja}}\ and\ \bibinfo {author} {\bibfnamefont {D.}~\bibnamefont
  {Loss}},\ }\bibfield  {title} {\enquote {\bibinfo {title} {Fermionic and
  {Majorana} bound states in hybrid nanowires with non-uniform spin-orbit
  interaction}}, }\href {https://doi.org/10.1140/epjb/e2015-50882-2} {\bibfield
   {journal} {\bibinfo  {journal} {Eur. Phys. J. B}\ }\textbf {\bibinfo
  {volume} {88}},\ \bibinfo {pages} {62} (\bibinfo {year} {2015})}\BibitemShut
  {NoStop}%
\bibitem [{\citenamefont {Dolcini}\ and\ \citenamefont
  {Rossi}(2018)}]{dolcini2018magnetic}%
  \BibitemOpen
  \bibfield  {author} {\bibinfo {author} {\bibfnamefont {F.}~\bibnamefont
  {Dolcini}}\ and\ \bibinfo {author} {\bibfnamefont {F.}~\bibnamefont
  {Rossi}},\ }\bibfield  {title} {\enquote {\bibinfo {title} {Magnetic field
  effects on a nanowire with inhomogeneous {Rashba} spin-orbit coupling: Spin
  properties at equilibrium}}, }\href
  {https://doi.org/10.1103/PhysRevB.98.045436} {\bibfield  {journal} {\bibinfo
  {journal} {Phys. Rev. B}\ }\textbf {\bibinfo {volume} {98}},\ \bibinfo
  {pages} {045436} (\bibinfo {year} {2018})}\BibitemShut {NoStop}%
\bibitem [{\citenamefont {Hansen}\ \emph {et~al.}(2018)\citenamefont {Hansen},
  \citenamefont {Danon},\ and\ \citenamefont {Flensberg}}]{hansen2018probing}%
  \BibitemOpen
  \bibfield  {author} {\bibinfo {author} {\bibfnamefont {E.~B.}\ \bibnamefont
  {Hansen}}, \bibinfo {author} {\bibfnamefont {J.}~\bibnamefont {Danon}}, \
  and\ \bibinfo {author} {\bibfnamefont {K.}~\bibnamefont {Flensberg}},\
  }\bibfield  {title} {\enquote {\bibinfo {title} {Probing electron-hole
  components of subgap states in {Coulomb} blockaded {Majorana} islands}},
  }\href {https://doi.org/10.1103/PhysRevB.97.041411} {\bibfield  {journal}
  {\bibinfo  {journal} {Phys. Rev. B}\ }\textbf {\bibinfo {volume} {97}},\
  \bibinfo {pages} {041411(R)} (\bibinfo {year} {2018})}\BibitemShut {NoStop}%
\bibitem [{Sup()}]{Supp}%
  \BibitemOpen
  \href@noop {} {\bibinfo  {journal} {See Supplemental Material for calculation
  details, which includes Refs. [27,31-36,50,59,60,64,71,72]}\ }\BibitemShut
  {NoStop}%
\bibitem [{\citenamefont {Stanescu}\ \emph {et~al.}(2013)\citenamefont
  {Stanescu}, \citenamefont {Lutchyn},\ and\ \citenamefont {{Das
  Sarma}}}]{stanescu2013dimensional}%
  \BibitemOpen
\bibfield  {journal} {  }\bibfield  {author} {\bibinfo {author} {\bibfnamefont
  {T.~D.}\ \bibnamefont {Stanescu}}, \bibinfo {author} {\bibfnamefont {R.~M.}\
  \bibnamefont {Lutchyn}}, \ and\ \bibinfo {author} {\bibfnamefont
  {S.}~\bibnamefont {{Das Sarma}}},\ }\bibfield  {title} {\enquote {\bibinfo
  {title} {Dimensional crossover in spin-orbit-coupled semiconductor nanowires
  with induced superconducting pairing}}, }\href
  {https://doi.org/10.1103/PhysRevB.87.094518} {\bibfield  {journal} {\bibinfo
  {journal} {Phys. Rev. B}\ }\textbf {\bibinfo {volume} {87}},\ \bibinfo
  {pages} {094518} (\bibinfo {year} {2013})}\BibitemShut {NoStop}%
\bibitem [{\citenamefont {Mishmash}\ \emph {et~al.}(2016)\citenamefont
  {Mishmash}, \citenamefont {Aasen}, \citenamefont {Higginbotham},\ and\
  \citenamefont {Alicea}}]{mishmash2016approaching}%
  \BibitemOpen
  \bibfield  {author} {\bibinfo {author} {\bibfnamefont {R.~V.}\ \bibnamefont
  {Mishmash}}, \bibinfo {author} {\bibfnamefont {D.}~\bibnamefont {Aasen}},
  \bibinfo {author} {\bibfnamefont {A.~P.}\ \bibnamefont {Higginbotham}}, \
  and\ \bibinfo {author} {\bibfnamefont {J.}~\bibnamefont {Alicea}},\
  }\bibfield  {title} {\enquote {\bibinfo {title} {Approaching a topological
  phase transition in {Majorana} nanowires}}, }\href
  {https://doi.org/10.1103/PhysRevB.93.245404} {\bibfield  {journal} {\bibinfo
  {journal} {Phys. Rev. B}\ }\textbf {\bibinfo {volume} {93}},\ \bibinfo
  {pages} {245404} (\bibinfo {year} {2016})}\BibitemShut {NoStop}%
\bibitem [{\citenamefont {Huang}\ \emph {et~al.}(2018)\citenamefont {Huang},
  \citenamefont {Pan}, \citenamefont {Liu}, \citenamefont {Sau}, \citenamefont
  {Stanescu},\ and\ \citenamefont {Das~Sarma}}]{huang2018etamorphosis}%
  \BibitemOpen
  \bibfield  {author} {\bibinfo {author} {\bibfnamefont {Y.}~\bibnamefont
  {Huang}}, \bibinfo {author} {\bibfnamefont {H.}~\bibnamefont {Pan}}, \bibinfo
  {author} {\bibfnamefont {C.-X.}\ \bibnamefont {Liu}}, \bibinfo {author}
  {\bibfnamefont {J.~D.}\ \bibnamefont {Sau}}, \bibinfo {author} {\bibfnamefont
  {T.~D.}\ \bibnamefont {Stanescu}}, \ and\ \bibinfo {author} {\bibfnamefont
  {S.}~\bibnamefont {Das~Sarma}},\ }\bibfield  {title} {\enquote {\bibinfo
  {title} {Metamorphosis of {Andreev} bound states into {Majorana} bound states
  in pristine nanowires}}, }\href
  {https://link.aps.org/doi/10.1103/PhysRevB.98.144511} {\bibfield  {journal}
  {\bibinfo  {journal} {Phys. Rev. B}\ }\textbf {\bibinfo {volume} {98}},\
  \bibinfo {pages} {144511} (\bibinfo {year} {2018})}\BibitemShut {NoStop}%
\bibitem [{\citenamefont {Vuik}\ \emph {et~al.}(2016)\citenamefont {Vuik},
  \citenamefont {Eeltink}, \citenamefont {Akhmerov},\ and\ \citenamefont
  {Wimmer}}]{vuik2016effects}%
  \BibitemOpen
  \bibfield  {author} {\bibinfo {author} {\bibfnamefont {A.}~\bibnamefont
  {Vuik}}, \bibinfo {author} {\bibfnamefont {D.}~\bibnamefont {Eeltink}},
  \bibinfo {author} {\bibfnamefont {A.}~\bibnamefont {Akhmerov}}, \ and\
  \bibinfo {author} {\bibfnamefont {M.}~\bibnamefont {Wimmer}},\ }\bibfield
  {title} {\enquote {\bibinfo {title} {Effects of the electrostatic environment
  on the {Majorana} nanowire devices}}, }\href
  {https://doi.org/10.1088/1367-2630/18/3/033013} {\bibfield  {journal}
  {\bibinfo  {journal} {New J. Phys.}\ }\textbf {\bibinfo {volume} {18}},\
  \bibinfo {pages} {033013} (\bibinfo {year} {2016})}\BibitemShut {NoStop}%
\bibitem [{\citenamefont {Woods}\ \emph {et~al.}(2018)\citenamefont {Woods},
  \citenamefont {Stanescu},\ and\ \citenamefont {{Das
  Sarma}}}]{woods2018effective}%
  \BibitemOpen
  \bibfield  {author} {\bibinfo {author} {\bibfnamefont {B.~D.}\ \bibnamefont
  {Woods}}, \bibinfo {author} {\bibfnamefont {T.~D.}\ \bibnamefont {Stanescu}},
  \ and\ \bibinfo {author} {\bibfnamefont {S.}~\bibnamefont {{Das Sarma}}},\
  }\bibfield  {title} {\enquote {\bibinfo {title} {Effective theory approach to
  the {Schr{\"o}dinger-Poisson} problem in semiconductor {Majorana} devices}},
  }\href {https://doi.org/10.1103/PhysRevB.98.035428} {\bibfield  {journal}
  {\bibinfo  {journal} {Phys. Rev. B}\ }\textbf {\bibinfo {volume} {98}},\
  \bibinfo {pages} {035428} (\bibinfo {year} {2018})}\BibitemShut {NoStop}%
\bibitem [{\citenamefont {Mikkelsen}\ \emph {et~al.}(2018)\citenamefont
  {Mikkelsen}, \citenamefont {Kotetes}, \citenamefont {Krogstrup},\ and\
  \citenamefont {Flensberg}}]{mikkelsen2018hybridization}%
  \BibitemOpen
  \bibfield  {author} {\bibinfo {author} {\bibfnamefont {A.~E.~G.}\
  \bibnamefont {Mikkelsen}}, \bibinfo {author} {\bibfnamefont {P.}~\bibnamefont
  {Kotetes}}, \bibinfo {author} {\bibfnamefont {P.}~\bibnamefont {Krogstrup}},
  \ and\ \bibinfo {author} {\bibfnamefont {K.}~\bibnamefont {Flensberg}},\
  }\bibfield  {title} {\enquote {\bibinfo {title} {Hybridization at
  superconductor-semiconductor interfaces}}, }\href {\doibase
  10.1103/PhysRevX.8.031040} {\bibfield  {journal} {\bibinfo  {journal} {Phys.
  Rev. X}\ }\textbf {\bibinfo {volume} {8}},\ \bibinfo {pages} {031040}
  (\bibinfo {year} {2018})}\BibitemShut {NoStop}%
\bibitem [{\citenamefont {Antipov}\ \emph {et~al.}(2018)\citenamefont
  {Antipov}, \citenamefont {Bargerbos}, \citenamefont {Winkler}, \citenamefont
  {Bauer}, \citenamefont {Rossi},\ and\ \citenamefont
  {Lutchyn}}]{antipov2018effects}%
  \BibitemOpen
  \bibfield  {author} {\bibinfo {author} {\bibfnamefont {A.~E.}\ \bibnamefont
  {Antipov}}, \bibinfo {author} {\bibfnamefont {A.}~\bibnamefont {Bargerbos}},
  \bibinfo {author} {\bibfnamefont {G.~W.}\ \bibnamefont {Winkler}}, \bibinfo
  {author} {\bibfnamefont {B.}~\bibnamefont {Bauer}}, \bibinfo {author}
  {\bibfnamefont {E.}~\bibnamefont {Rossi}}, \ and\ \bibinfo {author}
  {\bibfnamefont {R.~M.}\ \bibnamefont {Lutchyn}},\ }\bibfield  {title}
  {\enquote {\bibinfo {title} {Effects of gate-induced electric fields on
  semiconductor {Majorana} nanowires}}, }\href
  {https://doi.org/10.1103/PhysRevX.8.031041} {\bibfield  {journal} {\bibinfo
  {journal} {Phys. Rev. X}\ }\textbf {\bibinfo {volume} {8}},\ \bibinfo {pages}
  {031041} (\bibinfo {year} {2018})}\BibitemShut {NoStop}%
\bibitem [{\citenamefont {Stanescu}\ \emph {et~al.}(2011)\citenamefont
  {Stanescu}, \citenamefont {Lutchyn},\ and\ \citenamefont {{Das
  Sarma}}}]{stanescu2011majorana}%
  \BibitemOpen
  \bibfield  {author} {\bibinfo {author} {\bibfnamefont {T.~D.}\ \bibnamefont
  {Stanescu}}, \bibinfo {author} {\bibfnamefont {R.~M.}\ \bibnamefont
  {Lutchyn}}, \ and\ \bibinfo {author} {\bibfnamefont {S.}~\bibnamefont {{Das
  Sarma}}},\ }\bibfield  {title} {\enquote {\bibinfo {title} {{Majorana}
  fermions in semiconductor nanowires}}, }\href
  {https://doi.org/10.1103/PhysRevB.84.144522} {\bibfield  {journal} {\bibinfo
  {journal} {Phys. Rev. B}\ }\textbf {\bibinfo {volume} {84}},\ \bibinfo
  {pages} {144522} (\bibinfo {year} {2011})}\BibitemShut {NoStop}%
\bibitem [{\citenamefont {Reeg}\ \emph
  {et~al.}(2018{\natexlab{a}})\citenamefont {Reeg}, \citenamefont {Loss},\ and\
  \citenamefont {Klinovaja}}]{reeg2018metallization}%
  \BibitemOpen
  \bibfield  {author} {\bibinfo {author} {\bibfnamefont {C.}~\bibnamefont
  {Reeg}}, \bibinfo {author} {\bibfnamefont {D.}~\bibnamefont {Loss}}, \ and\
  \bibinfo {author} {\bibfnamefont {J.}~\bibnamefont {Klinovaja}},\ }\bibfield
  {title} {\enquote {\bibinfo {title} {Metallization of a {Rashba} wire by a
  superconducting layer in the strong-proximity regime}}, }\href
  {https://doi.org/10.1103/PhysRevB.97.165425} {\bibfield  {journal} {\bibinfo
  {journal} {Phys. Rev. B}\ }\textbf {\bibinfo {volume} {97}},\ \bibinfo
  {pages} {165425} (\bibinfo {year} {2018}{\natexlab{a}})}\BibitemShut
  {NoStop}%
\bibitem [{\citenamefont {Kells}\ \emph {et~al.}(2012)\citenamefont {Kells},
  \citenamefont {Meidan},\ and\ \citenamefont {Brouwer}}]{kells2012near}%
  \BibitemOpen
  \bibfield  {author} {\bibinfo {author} {\bibfnamefont {G.}~\bibnamefont
  {Kells}}, \bibinfo {author} {\bibfnamefont {D.}~\bibnamefont {Meidan}}, \
  and\ \bibinfo {author} {\bibfnamefont {P.~W.}\ \bibnamefont {Brouwer}},\
  }\bibfield  {title} {\enquote {\bibinfo {title} {Near-zero-energy end states
  in topologically trivial spin-orbit coupled superconducting nanowires with a
  smooth confinement}}, }\href {\doibase 10.1103/PhysRevB.86.100503} {\bibfield
   {journal} {\bibinfo  {journal} {Phys. Rev. B}\ }\textbf {\bibinfo {volume}
  {86}},\ \bibinfo {pages} {100503(R)} (\bibinfo {year} {2012})}\BibitemShut
  {NoStop}%
\bibitem [{\citenamefont {Stanescu}\ and\ \citenamefont
  {Tewari}(2013{\natexlab{b}})}]{stanescu2013disentangling}%
  \BibitemOpen
  \bibfield  {author} {\bibinfo {author} {\bibfnamefont {T.~D.}\ \bibnamefont
  {Stanescu}}\ and\ \bibinfo {author} {\bibfnamefont {S.}~\bibnamefont
  {Tewari}},\ }\bibfield  {title} {\enquote {\bibinfo {title} {Disentangling
  {Majorana} fermions from topologically trivial low-energy states in
  semiconductor {Majorana} wires}}, }\href {\doibase
  10.1103/PhysRevB.87.140504} {\bibfield  {journal} {\bibinfo  {journal} {Phys.
  Rev. B}\ }\textbf {\bibinfo {volume} {87}},\ \bibinfo {pages} {140504(R)}
  (\bibinfo {year} {2013}{\natexlab{b}})}\BibitemShut {NoStop}%
\bibitem [{\citenamefont {Liu}\ \emph {et~al.}(2017)\citenamefont {Liu},
  \citenamefont {Sau}, \citenamefont {Stanescu},\ and\ \citenamefont {{Das
  Sarma}}}]{liu2017andreev}%
  \BibitemOpen
  \bibfield  {author} {\bibinfo {author} {\bibfnamefont {C.-X.}\ \bibnamefont
  {Liu}}, \bibinfo {author} {\bibfnamefont {J.~D.}\ \bibnamefont {Sau}},
  \bibinfo {author} {\bibfnamefont {T.~D.}\ \bibnamefont {Stanescu}}, \ and\
  \bibinfo {author} {\bibfnamefont {S.}~\bibnamefont {{Das Sarma}}},\
  }\bibfield  {title} {\enquote {\bibinfo {title} {{Andreev} bound states
  versus {Majorana} bound states in quantum dot-nanowire-superconductor hybrid
  structures: Trivial versus topological zero-bias conductance peaks}}, }\href
  {https://doi.org/10.1103/PhysRevB.96.075161} {\bibfield  {journal} {\bibinfo
  {journal} {Phys. Rev. B}\ }\textbf {\bibinfo {volume} {96}},\ \bibinfo
  {pages} {075161} (\bibinfo {year} {2017})}\BibitemShut {NoStop}%
\bibitem [{\citenamefont {Moore}\ \emph {et~al.}(2018)\citenamefont {Moore},
  \citenamefont {Stanescu},\ and\ \citenamefont {Tewari}}]{moore2018two}%
  \BibitemOpen
  \bibfield  {author} {\bibinfo {author} {\bibfnamefont {C.}~\bibnamefont
  {Moore}}, \bibinfo {author} {\bibfnamefont {T.~D.}\ \bibnamefont {Stanescu}},
  \ and\ \bibinfo {author} {\bibfnamefont {S.}~\bibnamefont {Tewari}},\
  }\bibfield  {title} {\enquote {\bibinfo {title} {Two-terminal charge
  tunneling: Disentangling {Majorana} zero modes from partially separated
  {Andreev} bound states in semiconductor-superconductor heterostructures}},
  }\href {https://doi.org/10.1103/PhysRevB.97.165302} {\bibfield  {journal}
  {\bibinfo  {journal} {Phys. Rev. B}\ }\textbf {\bibinfo {volume} {97}},\
  \bibinfo {pages} {165302} (\bibinfo {year} {2018})}\BibitemShut {NoStop}%
\bibitem [{\citenamefont {Vuik}\ \emph {et~al.}(2018)\citenamefont {Vuik},
  \citenamefont {Nijholt}, \citenamefont {Akhmerov},\ and\ \citenamefont
  {Wimmer}}]{vuik2018reproducing}%
  \BibitemOpen
  \bibfield  {author} {\bibinfo {author} {\bibfnamefont {A.}~\bibnamefont
  {Vuik}}, \bibinfo {author} {\bibfnamefont {B.}~\bibnamefont {Nijholt}},
  \bibinfo {author} {\bibfnamefont {A.}~\bibnamefont {Akhmerov}}, \ and\
  \bibinfo {author} {\bibfnamefont {M.}~\bibnamefont {Wimmer}},\ }\bibfield
  {title} {\enquote {\bibinfo {title} {Reproducing topological properties with
  quasi-{Majorana} states}}, }\href {https://arxiv.org/abs/1806.02801}
  {\bibfield  {journal} {\bibinfo  {journal} {arXiv:1806.02801}\ } (\bibinfo
  {year} {2018})}\BibitemShut {NoStop}%
\bibitem [{\citenamefont {Kjaergaard}\ \emph {et~al.}(2012)\citenamefont
  {Kjaergaard}, \citenamefont {W{\"o}lms},\ and\ \citenamefont
  {Flensberg}}]{kjaergaard2012majorana}%
  \BibitemOpen
  \bibfield  {author} {\bibinfo {author} {\bibfnamefont {M.}~\bibnamefont
  {Kjaergaard}}, \bibinfo {author} {\bibfnamefont {K.}~\bibnamefont
  {W{\"o}lms}}, \ and\ \bibinfo {author} {\bibfnamefont {K.}~\bibnamefont
  {Flensberg}},\ }\bibfield  {title} {\enquote {\bibinfo {title} {{Majorana}
  fermions in superconducting nanowires without spin-orbit coupling}}, }\href
  {https://doi.org/10.1103/PhysRevB.85.020503} {\bibfield  {journal} {\bibinfo
  {journal} {Phys. Rev. B}\ }\textbf {\bibinfo {volume} {85}},\ \bibinfo
  {pages} {020503(R)} (\bibinfo {year} {2012})}\BibitemShut {NoStop}%
\bibitem [{\citenamefont {Fu}(2010)}]{fu2010electron}%
  \BibitemOpen
  \bibfield  {author} {\bibinfo {author} {\bibfnamefont {L.}~\bibnamefont
  {Fu}},\ }\bibfield  {title} {\enquote {\bibinfo {title} {Electron
  teleportation via {Majorana} bound states in a mesoscopic superconductor}},
  }\href {https://doi.org/10.1103/PhysRevLett.104.056402} {\bibfield  {journal}
  {\bibinfo  {journal} {Phys. Rev. Lett.}\ }\textbf {\bibinfo {volume} {104}},\
  \bibinfo {pages} {056402} (\bibinfo {year} {2010})}\BibitemShut {NoStop}%
\bibitem [{\citenamefont {H{\"u}tzen}\ \emph {et~al.}(2012)\citenamefont
  {H{\"u}tzen}, \citenamefont {Zazunov}, \citenamefont {Braunecker},
  \citenamefont {Yeyati},\ and\ \citenamefont {Egger}}]{hutzen2012majorana}%
  \BibitemOpen
  \bibfield  {author} {\bibinfo {author} {\bibfnamefont {R.}~\bibnamefont
  {H{\"u}tzen}}, \bibinfo {author} {\bibfnamefont {A.}~\bibnamefont {Zazunov}},
  \bibinfo {author} {\bibfnamefont {B.}~\bibnamefont {Braunecker}}, \bibinfo
  {author} {\bibfnamefont {A.~L.}\ \bibnamefont {Yeyati}}, \ and\ \bibinfo
  {author} {\bibfnamefont {R.}~\bibnamefont {Egger}},\ }\bibfield  {title}
  {\enquote {\bibinfo {title} {{Majorana} single-charge transistor}}, }\href
  {https://doi.org/10.1103/PhysRevLett.109.166403} {\bibfield  {journal}
  {\bibinfo  {journal} {Phys. Rev. Lett.}\ }\textbf {\bibinfo {volume} {109}},\
  \bibinfo {pages} {166403} (\bibinfo {year} {2012})}\BibitemShut {NoStop}%
\bibitem [{\citenamefont {Higginbotham}\ \emph {et~al.}(2015)\citenamefont
  {Higginbotham}, \citenamefont {Albrecht}, \citenamefont {Kir{\v{s}}anskas},
  \citenamefont {Chang}, \citenamefont {Kuemmeth}, \citenamefont {Krogstrup},
  \citenamefont {Jespersen}, \citenamefont {Nyg{\aa}rd}, \citenamefont
  {Flensberg},\ and\ \citenamefont {Marcus}}]{higginbotham2015parity}%
  \BibitemOpen
  \bibfield  {author} {\bibinfo {author} {\bibfnamefont {A.~P.}\ \bibnamefont
  {Higginbotham}}, \bibinfo {author} {\bibfnamefont {S.~M.}\ \bibnamefont
  {Albrecht}}, \bibinfo {author} {\bibfnamefont {G.}~\bibnamefont
  {Kir{\v{s}}anskas}}, \bibinfo {author} {\bibfnamefont {W.}~\bibnamefont
  {Chang}}, \bibinfo {author} {\bibfnamefont {F.}~\bibnamefont {Kuemmeth}},
  \bibinfo {author} {\bibfnamefont {P.}~\bibnamefont {Krogstrup}}, \bibinfo
  {author} {\bibfnamefont {T.~S.}\ \bibnamefont {Jespersen}}, \bibinfo {author}
  {\bibfnamefont {J.}~\bibnamefont {Nyg{\aa}rd}}, \bibinfo {author}
  {\bibfnamefont {K.}~\bibnamefont {Flensberg}}, \ and\ \bibinfo {author}
  {\bibfnamefont {C.~M.}\ \bibnamefont {Marcus}},\ }\bibfield  {title}
  {\enquote {\bibinfo {title} {Parity lifetime of bound states in a
  proximitized semiconductor nanowire}}, }\href
  {https://doi.org/10.1038/nphys3461} {\bibfield  {journal} {\bibinfo
  {journal} {Nat. Phys.}\ }\textbf {\bibinfo {volume} {11}},\ \bibinfo {pages}
  {1017} (\bibinfo {year} {2015})}\BibitemShut {NoStop}%
\bibitem [{\citenamefont {L{\"u}}\ \emph {et~al.}(2016)\citenamefont {L{\"u}},
  \citenamefont {Lu},\ and\ \citenamefont {Shen}}]{lu2016enhanced}%
  \BibitemOpen
  \bibfield  {author} {\bibinfo {author} {\bibfnamefont {H.-F.}\ \bibnamefont
  {L{\"u}}}, \bibinfo {author} {\bibfnamefont {H.-Z.}\ \bibnamefont {Lu}}, \
  and\ \bibinfo {author} {\bibfnamefont {S.-Q.}\ \bibnamefont {Shen}},\
  }\bibfield  {title} {\enquote {\bibinfo {title} {Enhanced current noise
  correlations in a {Coulomb}-{Majorana} device}}, }\href
  {https://doi.org/10.1103/PhysRevB.93.245418} {\bibfield  {journal} {\bibinfo
  {journal} {Phys. Rev. B}\ }\textbf {\bibinfo {volume} {93}},\ \bibinfo
  {pages} {245418} (\bibinfo {year} {2016})}\BibitemShut {NoStop}%
\bibitem [{\citenamefont {van Heck}\ \emph {et~al.}(2016)\citenamefont {van
  Heck}, \citenamefont {Lutchyn},\ and\ \citenamefont
  {Glazman}}]{van2016conductance}%
  \BibitemOpen
  \bibfield  {author} {\bibinfo {author} {\bibfnamefont {B.}~\bibnamefont {van
  Heck}}, \bibinfo {author} {\bibfnamefont {R.~M.}\ \bibnamefont {Lutchyn}}, \
  and\ \bibinfo {author} {\bibfnamefont {L.~I.}\ \bibnamefont {Glazman}},\
  }\bibfield  {title} {\enquote {\bibinfo {title} {Conductance of a
  proximitized nanowire in the {Coulomb} blockade regime}}, }\href {\doibase
  10.1103/PhysRevB.93.235431} {\bibfield  {journal} {\bibinfo  {journal} {Phys.
  Rev. B}\ }\textbf {\bibinfo {volume} {93}},\ \bibinfo {pages} {235431}
  (\bibinfo {year} {2016})}\BibitemShut {NoStop}%
\bibitem [{\citenamefont {Grabert}\ and\ \citenamefont
  {Devoret}(2013)}]{grabert2013single}%
  \BibitemOpen
  \bibfield  {author} {\bibinfo {author} {\bibfnamefont {H.}~\bibnamefont
  {Grabert}}\ and\ \bibinfo {author} {\bibfnamefont {M.~H.}\ \bibnamefont
  {Devoret}},\ }\href@noop {} {\emph {\bibinfo {title} {Single charge
  tunneling: {Coulomb} blockade phenomena in nanostructures}}}\ (\bibinfo
  {publisher} {Springer Science \& Business Media, New York},\ \bibinfo {year}
  {2013})\BibitemShut {NoStop}%
\bibitem [{\citenamefont {Reeg}\ \emph
  {et~al.}(2018{\natexlab{b}})\citenamefont {Reeg}, \citenamefont {Dmytruk},
  \citenamefont {Chevallier}, \citenamefont {Loss},\ and\ \citenamefont
  {Klinovaja}}]{Reeg18zero}%
  \BibitemOpen
  \bibfield  {author} {\bibinfo {author} {\bibfnamefont {C.}~\bibnamefont
  {Reeg}}, \bibinfo {author} {\bibfnamefont {O.}~\bibnamefont {Dmytruk}},
  \bibinfo {author} {\bibfnamefont {D.}~\bibnamefont {Chevallier}}, \bibinfo
  {author} {\bibfnamefont {D.}~\bibnamefont {Loss}}, \ and\ \bibinfo {author}
  {\bibfnamefont {J.}~\bibnamefont {Klinovaja}},\ }\bibfield  {title} {\enquote
  {\bibinfo {title} {Zero-energy {Andreev} bound states from quantum dots in
  proximitized {Rashba} nanowires}}, }\href {\doibase
  10.1103/PhysRevB.98.245407} {\bibfield  {journal} {\bibinfo  {journal} {Phys.
  Rev. B}\ }\textbf {\bibinfo {volume} {98}},\ \bibinfo {pages} {245407}
  (\bibinfo {year} {2018}{\natexlab{b}})}\BibitemShut {NoStop}%
\end{thebibliography}

%

\end{document}